\documentclass[graybox,natbib,nosecnum]{svmult}
\bibpunct{(}{)}{;}{a}{}{,} % suppress commas between author-names and year

\pdfoutput=1   %forces use of pdflatex. Disable if you prefer to use .eps or .ps figures.
\usepackage{mathptmx}       % selects Times Roman as basic font
\usepackage{helvet}         % selects Helvetica as sans-serif font
\usepackage{courier}        % selects Courier as typewriter font
\usepackage{type1cm}        % activate if the above 3 fonts are
\usepackage{makeidx}         % allows index generation
\usepackage{graphicx}        % standard LaTeX graphics tool
\usepackage{multicol}        % used for the two-column index
\usepackage[bottom]{footmisc}% places footnotes at page bottom
\usepackage[normalem]{ulem}	% for strike-through of text with \sout{}  
\usepackage[hidelinks,colorlinks=true,linkcolor=blue,citecolor=blue]{hyperref}  %for hyperlinks

\usepackage{amsmath}
\usepackage{amssymb}

\usepackage{soul}   % for high-lighting of text
\def\mearth{{\,M_\oplus}}
\def\rearth{{\,R_\oplus}}

\newcommand{\hbindex}[1]{\sethlcolor{white} \hl{#1}\index{#1}}  %highlights index entries

\makeindex             % used for the subject index
\begin{document}

\title*{Formation of Terrestrial Planets}
\author{Matthew S. Clement, Andr\'e  Izidoro , Sean N. Raymond \& Rogerio Deienno}
\authorrunning{Clement et al.} %for an abbreviated version of
\institute{Matthew S. Clement \at Johns Hopkins APL, 11100 Johns Hopkins Rd, Laurel, MD 20723, USA \email{matt.clement@jhuapl.edu} \and Andr\'e Izidoro \at Department of Earth, Environmental and Planetary Sciences, MS 126, Rice University, Houston, TX 77005, USA \and Sean N. Raymond \at Laboratoire d'Astrophysique de Bordeaux, Univ. Bordeaux, CNRS, B18N, allée Geoffroy Saint-Hilaire, 33615 Pessac, France \and Rogerio Deienno \at Southwest Research Institute, 1050 Walnut St. Suite 300, Boulder, CO 80302, USA}
\maketitle

\abstract{ Our understanding of the process of terrestrial planet formation has grown markedly over the past 20 years, yet key questions remain. This review begins by first addressing the critical, earliest stage of dust coagulation and concentration.  While classic studies revealed how objects that grow to $\sim$meter sizes are rapidly removed from \hbindex{protoplanetary disk}s via orbital decay (seemingly precluding growth to larger sizes), this chapter addresses how this is resolved in contemporary, \hbindex{streaming instability} models that favor rapid planetesimal formation via \hbindex{gravitational collapse} of solids in over-dense regions. Once formed, planetesimals grow into Mars-Earth-sized planetary embryos by a combination of pebble- and planetesimal accretion within the lifetime of the nebular disk. After the disk dissipates, these embryos typically experience a series of late giant impacts en route to attaining their final architectures. This review also highlights three different inner Solar System formation models that can match a number of empirical constraints, and also reviews ways that one or more might be ruled out in favor of another in the near future.  These include (1) the \hbindex{Grand Tack}, (2) the \hbindex{Early Instability} and (3) \hbindex{Planet Formation from Rings}.  Additionally, this chapter discusses formation models for the closest known analogs to our own terrestrial planets: super-Earths and terrestrial exoplanets in systems also hosting gas giants.  Finally, this review lays out a chain of events that may explain why the Solar System looks different than more than 99\% of exoplanet systems.
}

\section{Introduction}

Over the last three decades our understanding of the formation of terrestrial planets as grown markedly. While there is no universally accepted definition of a terrestrial planet, this class of worlds is commonly understood to include smaller (radii less than around $\sim$1.25 times that of the Earth) planets made mostly of silicate rocks and metals \citep[densities greater than $\sim$ 0.75 times that of the Earth:][]{luque22,margot24}. On the theoretical side, these advances were in large part catalyzed by interdisciplinary scientific efforts and technological advances (e.g., faster computers, better software).  Technology has also been crucial in driving observational advances as well. Although this field necessarily started within the context of our Solar System, new observational and theoretical studies have provided a push toward a more general, broadly-applicable framework.  While exoplanet science has undoubtedly revolutionized our knowledge of planet formation, the Solar System offers a fantastic wealth of well-characterised physical and chemical constraints that make it an unparalleled laboratory for refining and testing models of planet growth.
 
Radioactive dating and the determination of various Solar System bodies' chemical compositions have led to major advances. Constraints on the ages and compositions of different planets and small bodies directly connect with models of their origins and interiors.  Improvements in computational capabilities -- both in hardware and software -- have enabled more sophisticated and realistic numerical simulations that model a range of chemical and physical processes across all stages of planet formation. Modern planet formation theories are developed, tested and refined through interdisciplinary efforts leveraging empirical studies, geophysical and cosmochemical analyses, and dynamical simulations. Yet the formation of the terrestrial planets remains a subject of intense debate. At least three different models can explain the origins of our own inner Solar System. Each model is based on different fundamental assumptions, left to be disentangled by future research.

Low-mass, potentially terrestrial exoplanets are a hot topic in astronomy. The discovery of such planets has been a major success of planet-finding missions such as \hbindex{Kepler} ~\citep{boruckietal10} and \hbindex{TESS} \citep{ricker15}. The search for exo-terrestrial planets is especially exciting because they are potential candidates for hosting life as we know it. To date more than 5,700 exoplanets have been confirmed in over 4,000 different systems.  However, these observations have elucidated how the majority of planetary systems have dynamical architectures that are strikingly different from our own.  Gas giant exoplanets are relatively rare \citep[occuring in $\lesssim$10$\%$ of systems, depending on spectral type:][]{petigura18,gan22,beleznay22,bryant23}.  Moreover, they are often observed on orbits very different from those of Jupiter and Saturn; either very close to their central stars or on extremely eccentric orbits~\citep[e.g.][]{butleretal06,udryetal07}. Compact systems of \hbindex{hot super-Earth}s and \hbindex{sub-Neptune}s -- with sizes between 1 and 4 Earth radii; or masses between 1 and 20 Earth mass -- are systematically found orbiting their stars on orbits much smaller than that of Mercury~\citep{howardetal10,mayoretal11}.  While this class of planet seems to be present around the majority of main sequence stars~\citep[e.g.][]{howardetal12,fressinetal13,petiguraetal13,chiang13,zhu18}, no such planet exists in our inner Solar System. Given that most exoplanet systems look dramatically different than our own, it is not immediately obvious whether or not we expect those planets to have formed in the manner as our own terrestrial planets, or by completely different processes.

This article reviews terrestrial planet formation in the Solar System and around other stars. It discusses the dynamical processes that shaped the inner Solar System and expands on different formation pathways that can explain the orbital architectures of exoplanet systems.  Moreover, it presents a path toward understanding how our Solar System fits in the larger context of exoplanets, and how exoplanets themselves can be used to improve our understanding of our own Solar System.

Any successful model of Solar System formation must explain why Mars and Mercury are so much smaller than the neighboring Earth and Venus. Indeed, neighboring exoplanets tend to have very similar masses and radii \citep{millholland17}.  Additionally, the asteroid belt's striking low mass and dynamically excited state presents a key constraint for terrestrial planet formation models. This review focuses on three well-tested models that are able to consistently explain each of these peculiar inner Solar System traits. Likewise, any successful model for the formation of super-Earths must match the observed distributions or exoplanet masses and orbital periods. This chapter also discusses the relationships between super-Earths and true `terrestrial planets', and present formation models that demonstrate how the origins of the two classes of systems are likely far different than one might expect based solely on their observed sizes.  

%This review is organized as follows. It first briefly summarizes the early stages of planet formation that are responsible for turning dust into planetesimals. Next, we review the latter stages of terrestrial planet accretion in the Solar System. We then discuss the numerical tools used to study terrestrial planet formation, the ingredients for such simulations, and specific Solar System constraints on those simulations.  We subsequently present different explanations for the origin of the inner Solar System and possible ways to discriminate between the various models. Next we turn our attention toward terrestrial planet formation in the context of exoplanets. We discuss the orbital architectures of observed exoplanets, compositional constraints, the fate of terrestrial planets in systems with \hbindex{hot super-Earth}s and gas giant planets, and the dynamical evolution of exoplanet systems. Finally, we synthesize these different views and paint a picture of the Solar System's formation in the context of exoplanets. 

%Should we mention here previous reviews

\section{The early stages of planet formation}

Planet formation starts during star formation. Protostars begin to take shape when the densest parts of a molecular cloud of gas and dust collapses under \hbindex{self-gravity} \citep{shuetal1987,mckeeostriker07,andreetal14}. Conservation of angular momentum turns the surrounding clump around the forming star into a circumstellar disk, the birthplace of planets \cite[e.g.][]{safronov72}. The primary empirical evidence that planets form in this manner comes from the geometry of our Solar System.  The planets' orbits are almost perfectly coplanar and orbit the Sun in a common direction. The youngest \hbindex{protoplanetary disk}s are made up of around 99\% gas and 1\% dust grains or ice particles \cite[as is the case for the interstellar medium, e.g.][]{willianscieza11,bae23}. Despite the two orders of magnitude difference in abundance between these two constituents, until recently little was known about the gas component as it is much harder to observe than the dust since different gas species emit at specific wavelengths that require very high resolution spectroscopy to resolve \citep{najitaetal07,willianscieza11,dutreyetal04,oberg21}. For decades, the gas's solid counterpart has been used quite successfully as a disk tracer. Infrared and submillimeter observations of dust emission, in addition to optical/near-IR scattered light imagery with coronagraphs have helped constrain the distribution of dust in rotating, pressure supported disk-like structures \citep[e.g.][]{smithterrile84,beckwithetal90,odellwen94,hollandetal98,koerneretal98,schneideretal99}. Large far-infrared excesses have also been used to quantify how the thicknesses of disks increase radially \citep[also referred to as ``flaring:''][]{kenyonhartmann87}. Recent \hbindex{ALMA} observations have provided an unprecedented level of insight into planet forming-disks around young stars by revealing series of carved ring-type structures \citep{vandermareletal13,isellaetal13,almaetal15,nomuraetal16,fedeleetal17}. These features have been interpreted as signposts of planet formation in action \citep{dongetal15,bae19}, or dust responding to the gas via drag redistribution \citep[e.g.][]{takeuchiartymowicz01,flocketal15}.  In contrast,  molecular line emissions provides the best, and sometimes only, insight into the  density, chemistry, temperature, and kinematics of the gas~\citep[e.g.][]{pietuetal07, qietal11}. State of the art studies using \hbindex{ALMA} are currently only capable of resolving emission lines at tens of milliarcsecond-scales, and probing the gas disk properties at regions as close as $\gtrsim5-10$~au from the central star~\cite[e.g.][]{andrews20,oberg21}.

The mass-loss rate, thermal structure and lifetime of the disk all represent key inputs and constraints for planet formation models.  Observationally derived limits on these parameters are thus extremely valuable.  Thermal constraints, for example, can be teased out from the fact that dust grains re-radiate stellar photons at different wavelengths depending on their temperature \citep[e.g.][]{willianscieza11}. 
 
 Solids orbit their central star at the nominal Keplerian speed. Contrarily gases orbit at a slightly sub-Keplerian velocity because a radial gas pressure gradient partially supports \hbindex{protoplanetary disk}s against the central star's gravity. While the dust and gas components are well-mixed at the onset of disk formation, after just a short amount of time stellar gravity tends to force dust grains to settle into the disk mid-plane \citep[e.g.][]{weidenschilling80,nakagawaetal86}. Observations and simulations also suggest that the disk itself is in a constant state of infall towards the central stars as a consequence of radial angular momentum transport \citep[e.g.][]{papaloizoulin95,balbus03,dullemondetal07,armitage11}. As the disk loses mass onto the central star through this process \citep[and also through photo-evaporation by ultraviolet and X-ray radiation][]{gortihollenbach05}, it eventually transitions from optically thick to optically thin \citep{alexanderetal14}. Infra-red surveys of star-forming clusters have also been leveraged to place constraints on disk lifetimes \citep{haischetal01,hernandez07,mamajek09}.  Consistent with magnetospheric accretion models and stellar spectroscopy of gas accretion \citep{Hillenbrand08}, these surveys tend to conclude that the hot/inner parts of the disk do not live much longer than 10 Myr.  The lifetimes of \hbindex{protoplanetary disk}s is a fundamental constraint on planet formation. 

Within the framework of disk formation, evolution and eventual dissipation described above, it is convenient to divide planet formation into a series of key independent steps. Indeed, different physical and chemical processes are at play at different stages of planetary growth. For example, while the growth of micrometer-sized dust grains is driven by forces at the intermolecular level, the growth of $\gtrsim$km-sized bodies is dominated by gravity. In addition, the sheer number of dust particles involved in the planet formation process makes it impossible to use a single numerical simulation to study all phases of growth at all locations in the disk simultaneously.  Therefore, most studies tend to focus on a specific stage of the process. The following sections briefly address each of these stages independently.

\subsection{From dust to pebbles}

The solids available for planet formation are not uniformly distributed across a planet-forming disk.  At any given radius, only species with condensation/sublimation temperatures below the local temperature can exist as solids~\citep{grossmanlarimer74}. A condensation or sublimation front is a point in the disk, inside of which a given element or molecule can only be found in its gaseous form.  In planet formation models, the most commonly invoked fronts are the water \hbindex{snowline}, the CO \hbindex{snowline} and the silicate sublimation front \citep[e.g.][]{morbidellietal16,morby21,izidoro22_natast}.  Of course, as the disk cools, the locations of different condensation fronts sweep inward over time \citep[e.g.][]{lecaretal06,dodsonrobinsonetal09,hasegawapudritz11,martinlivio12}.  As a result of condensation and sublimation, iron and silicates are abundant in the inner regions of the disk, while the outer regions are rich in ice and other volatiles \citep{lodders03}. 

The first stage of planet formation involves the growth of millimeter and cm-sized dust aggregates from micrometer-sized dust and ice particles \citep[e.g.][]{lissauer93}. This view is well supported by disk observations and analyses of chondritic \hbindex{meteorites}.  Indeed, observations of young stellar objects at millimeter and centimeter wavelengths detect dust grains at these size ranges \citep{testietal03,wilneretal05,rodmannetal06,braueretal07,nattaetal07,riccietal10,testietal14,ansdeelletal17}, and wave polarization in these observations has been interpreted to suggest that dust aggregates are relatively compact \citep{tazaki19,kirchschlager19,brunngraberandwolf21}.  Likewise, primitive \hbindex{meteorites} are mainly composed of millimeter-sized silicate spheres known as chondrules \citep{shuetal01,scott07} that are cemented together within a fine-grained matrix of calcium aluminum rich inclusions (\hbindex{CAIs}), unprocessed interstellar grains and primitive organics \citep{scotttaylor83,scottkrot14}. Given a U-Pb chronometer-derived age estimate of 4.567 Gyr \citep{amelin10,connellyetal12}, \hbindex{CAIs} are thought to have been the first solids to condense in the Sun's \hbindex{protoplanetary disk} \citep[e.g.][]{bouvierwadhwa10,dauphaschaussidon11}. Chondrule formation likely started around the same time \hbindex{CAIs} were forming and continued for a few million years \citep{connellyetal12}.  It is also possible, that chondrules are actually the result of impact jets produced during planetesimal collisions \citep[e.g.][]{asphaugetal11,johnsonetal15,wakita17,lichtenbergetal17}.  However, most studies conclude that their properties are more consistent with being rapidly-heated aggregates of mm-sized silicate dust grains \citep{deschconnolly02,morrisdesch10}.

Laboratory experiments and numerical simulations also tend to conclude that the first stage of dust and ice growth is dominated by hit-and-stick collisions \citep[][]{blumwurm00,blumetal00,poppeetal00,guttleretal10,testietal14}.  In this early stage of accretion, Electrostatic charges, magnetic material effects \citep{dominiknubold02,okuzumi09} and adhesive van der Wals forces  \citep{heimetal99,gundlachetal11} largely determine the probability that two dust particles will stick and stay joined together. Through these processes micrometer sized dust grains stick together and form larger fluffy porous aggregates \citep{blumwurm08}, and are eventually compacted by collisions to mm or cm-sized grains \citep{ormeletal07,zsometal10,schrapler22}. There is a general consensus that collisional growth of micrometer-sized dust grains is efficient up to mm to cm range in sufficiently dense regions of \hbindex{protoplanetary disk}s.

\subsection{From pebbles to Planetesimals}

Growth beyond cm-sized aggregates faces several obstacles \citep[e.g][]{weidenschilling77,testietal14}.  This subject is one of the most active current research areas in planet formation.
 
Numerical and laboratory experiments suggest that colliding mm and cm-sized dust grains do not grow up to meter and kilometer sized bodies \citep{chokshietal93,dominiketielens97,gortietal15,krijtetal16}. Depending on particles' sizes and impact velocities, colliding dust particles and/or aggregates may bounce off of each other instead of growing \citep[e.g.][]{wadaetal09}.  This is known as the ``\hbindex{bouncing barrier}''  \citep[e.g.][]{zsometal10,birnstieletal11,testietal14}.  Because the orbital speed of nebular gas is slightly sub-Keplerian, solid particles on a Keplerian orbits essentially ``feel'' a headwind that can be described by a drag force \citep{whipple72,adachietal76,weidenschilling77,haghighipourboss03}.  This force causes their orbits to decay, and spiral inwards towards the central star in very short timescales. Only sufficiently small particles that are strongly coupled to the gas do not drift as consequence of this aerodynamic effect. In contrast, decimeter to meter-sized particles lie in a regime where they feel a very strong relative drag force, and spiral inwards much quicker than $\gtrsim$km-size objects. A 1-meter size object at $\sim$1 AU in a typical disk falls toward the star in $\sim 100$ years \citep[e.g][]{weidenschilling77}. The large differences in the radial speeds of inwardly migrating particles with disparate sizes thus result in high-speed collisions; leading to bouncing, fragmentation or erosion \citep{krijetetal15} in non-turbulent disks.  This problem is known as the  drift-fragmentation ``barrier''.  In turbulent disks collisional velocities are regulated by turbulence, which in turn determines the physical size of the fragmentation barrier \citep{ormelandcuzzi07}.
%While the \hbindex{bouncing barrier} significantly limits the growth of similar-size dust grains colliding at energetic speeds, it may not represent the end of dust growth. Energetic impacts between small projectiles and relatively larger targets can lead to mass transfer, helping to alleviate the \hbindex{bouncing barrier} problem \citep{wurmetal05,teiserwurm09,kotheetal10,windmarketal12a,windmarketal12b}. However, even if mm or cm-size bodies can grow via pair-wise collisions \citep[e.g.][]{okuzumietal12}, growth towards the km-size scale is hindered by a second barrier.

The most likely way that nature overcomes these barriers and produces macroscopic solid bodies is by bypassing the critical size for rapid particle drift and directly forming planetesimals via \hbindex{gravitational collapse} \citep{goldreichward73,youdinshu02,johansenetal14,lesur23}. In order for mm or cm size particles to collapse they must be highly concentrated in a particular region of the disk. As discussed previously, dust particles are expected to sediment towards the disk mid-plane.  However, the collapse of a dense mid-plane layer is generally prevented by turbulent diffusion \citep{weidenschilling80,cuzzietal08,johansenetal09}. Different mechanisms have been proposed to operate with efficiencies sufficient to concentrate particles in local regions of the gas disk and trigger direct collapse \citep[see for example:][]{balbushawley91,kretkelin07,lyraetal08a,braueretal08,lyraetal08b,lyraetal09,chambers10,drazkowskaetal13,squirehopkins17,lichtenberg21,morby21,izidoro22_natast}.
%\hbindex{Gravitational collapse} only takes place if the local density is high enough to satisfy the following condition: $ 1 > (9 \Omega_c^2)/(4 \pi \rho_c G) $, where $\Omega_c$ is the Keplerian angular velocity at the concentrated clump location, G is the gravitational constant and $\rho_c$ is the mean bulk density of the potentially collapsing clump \citep[e.g.][]{binneytremaine08,reinetal10,johansenetal11,shichiang13}.  This expression is derived from the Toomre stability criterion for \hbindex{gravitational collapse} \citep{toomre64}; which quantifies the interplay between the stabilizing rotation,  de-stabilizing \hbindex{self-gravity}, and stabilizing  pressure of the clump \citep{binneytremaine08}. 

Turbulent motion of the gas can also concentrate particles in rotating substructures within the disk known as eddies \citep{cuzziweidenschilling06,chambers10}. Instabilities such as the vertical shear instability \citep{nelsonetal13,linyoudin15,barkerlatter15,umurhanetal16} and the baroclinic instability \citep{lyraklahr11,raettigetal13,bargeetal16,stollkley16} are also potential candidates for creating overly dense regions in the disk \citep{johansenlambrechts17}. Similarly, localized high pressure regions \citep[pressure bumps:][]{weidenschilling80,guilera20,chambers21} can trap and concentrate drifting particles and accelerate those in the vicinity of positive pressure gradients towards super Keplerian speeds. Through this process, inward drifting particles slow down, halt, or even begin to migrate outward depending on the steepness of the local pressure gradient \cite[e.g.][]{haghighipourboss03}.  Pressure bumps have been proposed to be consistent with high-resolution disk observations \citep{andrews18,dullemond18}, and may exist as the result of a sharp transition in the disc viscosity or due to tidal perturbations from a sufficiently large planet \citep[for a more detailed discussion see reviews by][]{chiangyoudin10,johansenetal14,johansenlambrechts17}. If the local density in solids increases by an order of magnitude at the pressure bump, the level of turbulence may be significantly counterbalanced, thus allowing for planetesimal formation by \hbindex{gravitational collapse} \citep{youdinshu02}. It is also possible for the local density of particles to be significantly increased via the delivery of slowly drifting small particles that are released as larger bodies fragment and sublimate when they encounter hotter regions of the disk \citep{sirono11,idaguillot16}. This may well be the case for mm or cm-sized pebbles that form in the water and volatiles-rich outer regions of the disk and drift inwards. Similarly, if inward drifting pebbles cross the water-ice \hbindex{snowline}, their water-content sublimates and small solid silicate/metal grains from within their interiors are released \citep{morbidellietal15}. 

The back-reaction friction force between mm or cm-sized particles and the gas can also serve as a mechanism for aerodynamically concentrating particles \citep{youdingoodman05}. Because the gas is orbiting at a sub-Keplerian velocity, the back reaction of \hbindex{gas drag} felt by a solid particle tends to accelerate the gas. If mm or cm-sized solid particles cluster sufficiently through this process, their collective back reaction becomes more pronounced.  This allows individual particles from the outer regions of the disk drifting at nominal speeds to join a more slowly drifting cluster in the process of forming \citep{johansenyoudin07}. If the local dust/gas ratio subsequently meets the threshold for collapse, the cluster eventually shrinks and leads to the rapid formation of planetesimals.  A more detailed physical explanation of the process can be found in \citet{magnan24}.  This process has been demonstrated to successful form planetesimals with sizes ranging from $\sim$1 to $\sim$1000 Km \citep{johansenetal07,baistone10,simonetal16,schaferetal17,carreraetal17,yang21}, and is typically referred to as the ``\hbindex{streaming instability}''. Streaming instabilities may be preferentially triggered outside the disk \hbindex{snowline} where sticky ice particles form \citep{drazkowskadullemond14,armitageetal16,drazkowskaalibert17}.  Although many questions remain (see chapters by Klahr and Armitage), this scenario represents the current consensus model for planetesimal formation.

\subsection{From planetesimals to planetary embryos}

Mutual gravitational interactions between large bodies plays an increasingly significant role as planetesimals continue to grow past the $\sim$1-100 km size range. At this stage planetesimals can grow larger by colliding and merging with other planetesimals, or by accreting the remaining mm or cm-sized grains in the gas disk. 

\subsubsection{Planetesimal-planetesimal growth:} 

Let us assume that there exists a population of planetesimals with $N$ members and total mass $Nm$ at 1 AU.  These planetesimals have formed by a combination of the mechanisms described above, and are still embedded in the gaseous \hbindex{protoplanetary disk}. The initial masses and physical radii of individual planetesimals are denoted $m$ and $R$, respectively.  This population of planetesimals will evolve through a series of collisions and gravitational scattering events. Close encounters between planetesimals increase their random velocities by increasing their orbital inclinations and eccentricities. The random velocity $v_{rnd}$ represents the deviation of the planetesimal's velocity from that of the Keplerian circular and planar orbit at its location:
\begin{equation}
v_{rnd} = (\frac{5}{8}e^2 + i^2)^{1/2} v_{k},
\end{equation}
where $e$  and $i$ are the planetesimal's orbital eccentricities and inclinations \citep{safronov72,greeenbergetal91}. $v_k=\sqrt{\frac{GM\odot}{r}}$ is the planetesimal's Keplerian velocity, where $G$ is the gravitational constant, $M_\odot$ is the mass of the central star, and $r$ is the planetesimal's orbital radius. 

Planetesimal growth rates are largely governed by their relative relocity with respect to one another. Of greatest importance are their random velocities, as well as the differential shear across the  \hbindex{Hill radii} of interacting bodies. The Hill radius $R_H$ of a planetesimal orbiting a star with mass $M_\odot$ is defined as
\begin{equation}
R_{H} = a\left( \frac{m}{3M_\odot}\right)^{1/3},
\end{equation}
where $a$ and $m$ are the planetesimal's semi-major axis and mass, respectively. 

Gravitational scattering events between planetesimals are in the shear-dominated regime when $v_{rnd} < R_H \Omega$, and in the dispersion-dominated regime when $v_{rnd} > R_H \Omega$, where $\Omega$ is the local Keplerian frequency. In the shear-dominated regime, planetesimals spend a significant amount of time in close proximity to one another during a close encounter. In this case, their orbits can be gravitationally deflected fairly significantly such that their trajectories are ``focused'' towards each other. This increases the probability of a collision occurring, thus speeding up their growth.  In the dispersion-dominated regime, however, interacting planetesimals spend much less time close to each other because of their high relative speeds. Therefore, gravitational focusing is less efficient, collisional probabilities decrease, and accretion timescales increase. 
 
A number of dissipative effects also conspire to damp the random velocities of planetesimals. Among others, these include physical collisions \citep{goldreichtremaine78}, \hbindex{gas drag} \citep{adachietal76} and gas \hbindex{dynamical friction} \citep{grishinperets15}. The balance between processes that 
 have tend to excite and those that damp orbits changes over time as the disk dissipates and planetesimals grow.

Characteristic 1-100 km-size planetesimals \citep{morbidellietal09,johansenetal12,delbo17} accrete other large bodies and become Moon-Mars-mass planetary embryos and eventually planets in three different growth regimes. In the ``\hbindex{runaway growth}'' regime, an initial generation of planet embryos form as the result of the fact that larger planetesimals grow faster than smaller ones.   In the ``\hbindex{orderly growth}'' regime large embryos grow at roughly the same rate, eventually forming planets. The ``\hbindex{oligarchic growth}'' regime is an intermediate growth regime between these two scenarios. 

One can write the accretion rate of a planetesimal with mass $m$ as \citep{safronov72,idanakazawa89,greenzweiglissauer90,rafikov03}
\begin{equation}
\frac{dm}{dt} \simeq \pi {R}^2 \Omega \Sigma \frac{v_{rnd}}{v_{rnd,z}} \left(1+\frac{2Gm}{R v_{rnd}^2} \right),
\end{equation}
where  $R$ is the planetesimal's physical radius, $\Omega$ is its Keplerian frequency, $\Sigma$ is the local planetesimal surface density, and $v_{rnd,z} \neq 0$ is the averaged planetesimal velocity's vertical component.  The term $\frac{2Gm}{R v_{rnd}^2}$ accounts for gravitational focusing and is simply the ratio of the escape velocity from the planetesimal surface to the relative velocity of the interacting planetesimals squared. $\Sigma \approx N  \overline{m} H$ quantifies the disk's vertical thickness (here N is the surface number density of planetesimals, $\overline{m}$ is the average mass of planetesimals and $H \approx v_{rnd,z}/\Omega$). Note, however, that Eq. 3 neglects bouncing, fragmentation or erosion during collisions.

A phenomena often referred to as ``\hbindex{dynamical friction}'' plays a crucial role in during the early phases of planetesimal accretion. As a result of angular momentum transfer during close-encounters with smaller bodies, the largest planetesimals in a given region of the disk tend to have their orbital eccentricities and inclinations damped. Through this process, their random velocities decrease; thus further increasing the rate at which they can accrete material through gravitational focusing. If (1) gravitational focusing is large (or planetesimal are small); (2) planetesimals' random velocities are roughly independent of their masses ($v_{rnd}$ and $v_{rnd,z}$); and (3) growth does not strongly affect the surface density of planetesimals, Equation 3 may be simplified as \citep{wetherillstewart89,kokuboida96}:
\begin{equation}
\frac{1}{m}\frac{dm}{dt} \approx t_{grow,run} \approx  \Sigma \frac{1}{v_{rnd,z}}\frac{{m}^{1/3}}{v_{rnd}} \propto {m}^{1/3}.
\end{equation}
In this regime the accretion rate of the most massive planetesimals tends to increase with time.  This is the so called ``\hbindex{runaway growth}'' regime of planetesimal accretion.
%If $m_1$ and $m_2$ are the masses of two planetesimals in the disk where $m_1 > m_2$, then $\frac{d(m_1/m_2)}{dt} > 0$.

%Of course, non-accretionary collisions also occur during the \hbindex{runaway growth} phase \citep{leinhardtrichardson05b,stewartleinhardt09}. If small fragments are produced during imperfect accretion events, they may have a second chance of being accreted by other planetesimals if the gaseous disk has not yet dissipated. Small fragments' orbits are quickly damped by \hbindex{gas drag}, which in turn increases the probably that they are consumed by larger planetesimals \citep{wetherillstewart93,rafikov04}. However, this is only possible if small fragments do not drift inwards too quickly to be accreted \citep{kenyonluu99,inabaetal03,kobayashietal10}.

An initial population of planetary embryos is the result of runaway accretion \citep{kokuboida96,ormeletal10}. At some point, gravitational perturbations between the emerging embryos have a larger effect on their dynamical evolution than interactions with smaller planetesimals, and their growth regime changes. The presence of sufficiently massive embryos increases the random velocities of planetesimals and significantly alters their local surface density \citep{tanakaida97}. Thus, the local velocity dispersion depends strongly on the mass of the largest embryo. As a result, the growth rate takes the form
\citep{wetherillstewart89,kokuboida98}:
\begin{equation}
\frac{1}{m}\frac{dm}{dt} \approx  t_{grow,orl} \propto  \Sigma \frac{1}{v_{rnd,z}}\frac{{m}^{-1/3}}{v_{rnd}} \propto m^{-1/3}
\end{equation}
This so-called``Oligarcic'' growth rate depends on the scaling of $\Sigma$, $v_{rnd}$ and $v_{rnd,z}$ \citep{rafikov03}. In this regime, planetary embryos still grow faster than planetesimals, however, small planetary embryos can grow faster than larger ones.  The growth of oligarchs is further accelerated in the vicinity of MMRs as a result of planetesimals concentrating and growing rapidly to intermediate masses in nearby first order resonances \citep{wallace19,wallace21}. According to \citep{idamakino93} the transition between the runaway and oligarchic regimes occurs when embryo masses are only a small fraction (a percent or so) of the total mass carried by the planetesimal population \citep[but see ][for a different criterion]{ormeletal10b}.

The ultimate result of the oligarchic phase is a bi-modal population of planetary embryos and planetesimals. At this phase, the total mass in embryos is still much smaller than the total mass in planetesimals. Planetary embryos growing in the thick sea of planetesimals are typically separated from one another by 5-10 mutual \hbindex{Hill radii} \citep{kokuboida95,kokuboida98,kokuboida00,ormeletal10}, where the mutual Hill radius of two embryos with masses $m_i$ and $m_j$ and semi-major axes $a_i$ and $a_j$ is defined as
\begin{equation}
R_{H,i,j} = \frac{a_i+a_j}{2}\left( \frac{m_i+m_j}{3M_\odot}\right)^{1/3}.
\end{equation}
If the spacing between neighboring embryos decreases to less than a few mutual \hbindex{Hill radii}, they scatter off of each other, thus bringing their orbital separation back to closer to $\sim$5$R_{H,i,j}$. After embryo-embryo scattering events, \hbindex{dynamical friction} (or \hbindex{gas drag}) tends to re-circularize their orbits and damp  their orbital inclinations provided that there is a sufficient quantity of mass in planetesimals or gas in the embryo's vicinity. 

\begin{figure}
% For example, with the graphicx style use
\centering
\includegraphics[scale=.15]{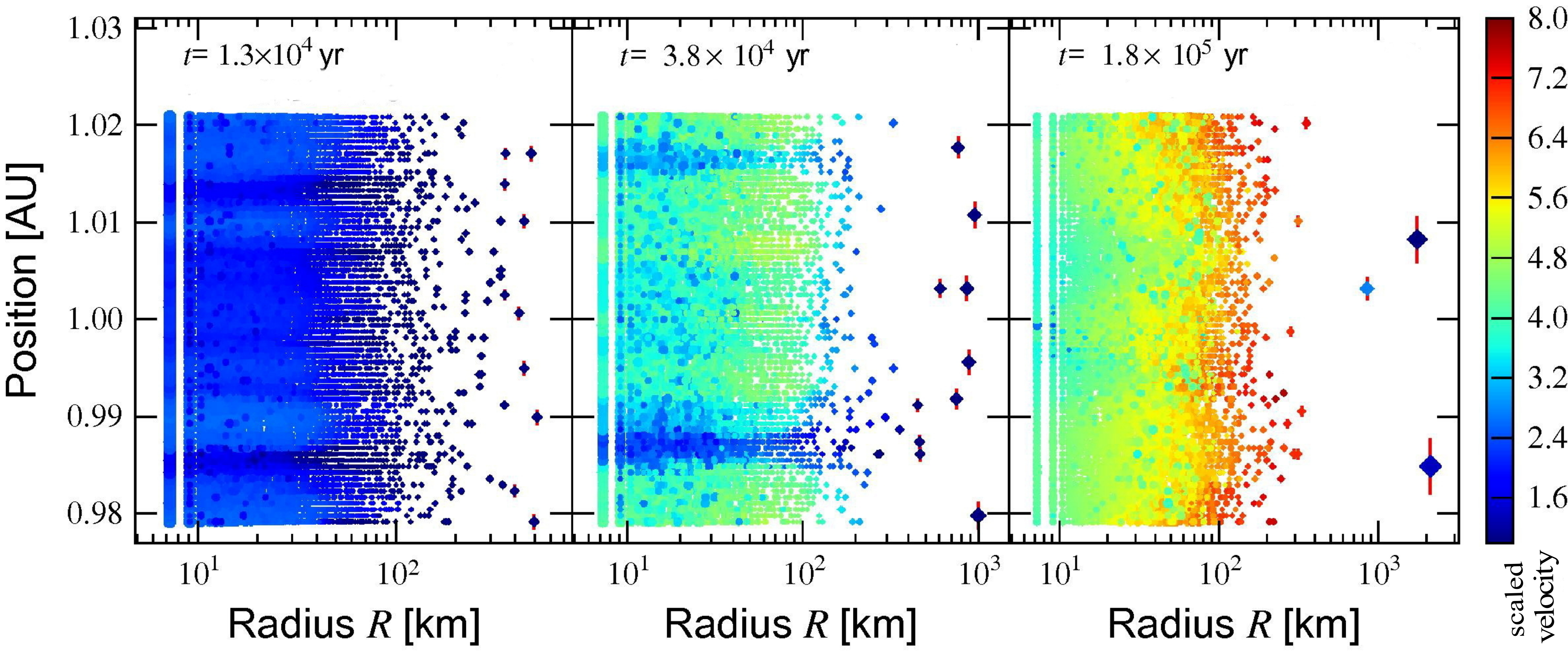}
\caption{Planetesimal and embryo growth in the oligarchic regime. Dots represent swarms of planetesimals which are simulated as single entity. The dot size scales with the total mass of the swarm. Individual large bodies are denoted with diamonds. The different colors represent the scaled random velocities of the bodies. The red bar intercepting the largest bodies in the system represents the respective  size of their Hill radius. Figure adapted from \citep{ormeletal10}}
\label{fig:1}       % Give a unique label
\end{figure}

Figure \ref{fig:1} shows the growth of planetesimals and embryos during the \hbindex{oligarchic growth} regime \citep{ormeletal10}. Three snapshots in the system's evolution are shown. The particles are modeled using a Monte Carlo method that computes the collisional and dynamical evolution of the system  \citep{ormeletal10}, and tracer particles to represent the much larger swarm of small planetesimals.  After 0.18 Myr, two prominent planetary embryos with radii larger than 1000 km emerge with a mutual separation of a few \hbindex{Hill radii}.

\subsubsection{Pebble Accretion: from planetesimals to planets}

If planetesimals form early they may accrete dust grains and aggregates drifting inward within the disk towards the central star. The existence of such grains in planet-forming disks has been observationally confirmed \citep[e.g.][]{testietal14}. The accretion of mm or cm-sized grains by a more massive body is commonly referred to as ``\hbindex{pebble accretion}'' \citep{johansenlacerda10,ormelklahr10,lambrechtsjohansen12,morbidellinesvorny12,johansenetal15,xuetal17}. 

Given appropriate conditions within the gas disk, \hbindex{pebble accretion} can be much faster than planetesimal accretion.  This may solve multiple long-standing problems, especially ones related to the formation of exoplanets and the solar system's giant planets \citep[for a recent review see][and references therein]{drazkowska23}.  In the classic, ``core accretion'' scenario for giant planet formation, gas giants form when their cores reach the threshold for runaway gas accretion \citep[$\sim$10$M_\oplus$,][]{mizuno80,pollacketal96}.  While this model axiomatically requires a core to form before the gaseous disk disperses, it is unclear whether the process of planetesimal accretion alone is sufficient to grow the cores of Jupiter or Saturn within a typical disk lifetime~\citep[e.g.][see also chapter by Youdin]{thommesetal03,levisonetal10}.

As discussed earlier in this chapter, millimeter- to centimeter-sized pebbles orbiting in gas disks rapidly spiral inwards due to \hbindex{gas drag} \citep{adachietal76,johansenetal15}. The dynamical behavior of a single drifting pebble approaching a planetesimal from a more distant orbit is determined by a competition between \hbindex{gas drag} and its gravitational interactions with the larger body. Assuming that the planetesimal is sufficiently small such that it does not disturb the background gas disk structure (e.g. gas disk velocity and density), two end states are possible. The pebble may either cross the planetesimal orbit without being accreted, or its original orbit can be sufficiently deflected to allow for accretion.

\hbindex{pebble accretion} can allow embryos and proto-planets to grow rapidly. However, planetesimals (or even planetary embryos) cannot grow indefinitely, even if the pebble flux is high. As an embryo grows, it gravitationally perturbs the structure of the gaseous disk. Eventually, the growing body opens a shallow gap in disk and creates a local pressure bump outside of its orbit. If the pressure bump is large enough, particles entering the bump are accelerated by the concentrated gas and stop drifting inwards. At the ``Pebble isolation mass,'' $M_{iso}$, an embryo or planet stops accreting pebbles,
\begin{equation}
M_{iso} = 20\left( \frac{H_{gas}/a_p}{0.05}\right)^3 M_\oplus,
\end{equation}
where $H_{gas}$ is the gas disk scale height \citep{lambrechtsetal14,morbidellinesvorny12}. It is worth noting that several studies have proposed that pebbles may be partially or even fully evaporated/destroyed before they can reach the accreting core. This effect may become important before the core reaches isolation mass \citep{alibert17,brouwersetal17}. Further study is needed to understand exactly how this effect limits embryo growth by \hbindex{pebble accretion}.

\subsection{From planetary embryos to planets} 

The final stage of terrestrial planet growth is thought to occur after the \hbindex{protoplanetary disk}'s gas dissipates; thus removing the dissipative mechanisms of \hbindex{gas drag} and gas \hbindex{dynamical friction}. In such an environment, gravitational focusing becomes negligible, and accretion timescales increase precipitously. Given that $v_{rnd} \approx v_{rnd,z}$ \citep{rafikov03} Eq 3 takes the form
\begin{equation}
\frac{1}{m}\frac{dm}{dt} \approx t_{grow,ord} \approx  \Sigma \frac{v_{rnd}}{v_{rnd,z}}{m}^{-1/3} \approx \Sigma {m}^{-1/3}
\end{equation}
In this growth regime -- termed  ``\hbindex{orderly growth}'' or ``late stage accretion'' --  $\Sigma$ decreases markedly with time as massive embryos accrete or scatter nearby planetesimals and open large gaps in the disk \citep{tanakaida97}. 
 This stage is marked by violent giant collisions between planetary embryos, power scattering events, and ejections of macroscopic bodies.  Therefore, the system's evolution is chaotic, and the total planetesimal population decreases drastically. Assuming that 50\% of the total mass in planetesimals is carried by embryos~\citep{kenyonbromley06}, the  mass of an embryo at the start of \hbindex{orderly growth} is $M_{ord}  = \int_{r-\Delta_r/2}^{r+\Delta r/2} 2 \pi r'  \Sigma(r')/2  dr' \simeq \pi r \Delta r \Sigma$, where  $\Delta r$ corresponds to the width of the feeding zone of the embryo  and r is the  planetary embryo's heliocentric distance \citep{lissauer87}. The size of the feeding zone of an embryo typically ranges between a few to 10$R_{H}$. 
%At this stage, most of the mass is carried by embryos rather than planetesimals. All planetary embryos grow at a similar rate and their mass ratios tend towards unity. If $m_1$ and $m_2$ are the masses of two planetary embryos growing in the orderly regime, where $m_1 > m2$, then $\frac{d(m_1/m_2)}{dt} \approx 0$.

%This kind o comes out of no where so I'm removing it
%In our Solar System, embryos probably reached  masses between those of the moon and Mars in the terrestrial region. This is consistent with the typical masses of planetary embryos produced in  oligarchic regime \citep{kokuboida00,chambers01}. The \hbindex{oligarchic growth} timescale of a Mars-mass embryo with bulk density $\rho_p=$ 3$g/cm^3$ orbiting at $a_p=1 AU$ and interacting with planetesimals with average random velocities  $v_{rnd}=v_{rnd,z}=0.02v_k$ is about 0.37 Myr for a local surface density in planetesimals of $\Sigma=10 g/cm^2$. The growth timescale of this same planetary embryo in the orderly regime  is about two orders of magnitude longer (23 Myr). This highlights the dramatic role of gravitational focusing.

\subsection{Methods and Numerical tools}

A number of techniques are utilized to constrain the nature of the initial planeteismal population in the inner Solar System (i.e. its chemical, radial mass and size frequency distributions).  Two and three dimensional hydrodynamical calculations including \hbindex{self-gravity} are ubiquitously used in computational studies of the \hbindex{streaming instability} \citep{youdingoodman05,johansenyoudin07,johansenetal09,baistone10,johansenetal12,simonetal16}.  The results of these models are then parameterized, and utilized in one dimensional alpha disk models that track dust growth, disk chemistry and planetesimal formation with the aim of extracting the authentic state of the terrestrial disk around the onset of \hbindex{runaway growth} \citep{birnstiel16,drazkowska18,charnoz19,appelgren20,lichtenberg21,morby21,izidoro22_natast}.

\subsubsection{Embryo and Planetesimal accretion}

Studies of the runaway and oligarchic regimes take a wide range of forms and leverage a range of computational methodologies. These includes \hbindex{N-body simulation}s \citep{idamakino93,aarsethetal93, kokuboida96,kokuboida98,kokuboida00,richardsonetal00,thommesetal03,barnesetal09,clement20_embryo,woo21_embryo}, analytical/semi-analytical calculations based on statistical algorithms \citep{greenbergetal78,wetherillstewart89,rafikov03,rafikov03b,rafikov03c,goldreichetal04,kenybromley04,idalin04,chambers06,morbidellietal09,schlichtingsari11,schlichtingetal13}, hybrid statistical/N-body (or N-body coagulation) codes which incorporates the two latter approaches \citep{spauteetal91,weidenschillingetal97,ormeletal10,bromleykenyon11,glaschkeetal14}, and finally the more recently developed hybrid particle-based algorithms \citep{levisonetal12,morishimaetal15,morishima17}.  Each tool is most optimized for modeling different specific stages of planet formation. While studies of the earlier epochs of planet formation are mostly conducted using analytical and statistical tools, the intermediate and late stages of accretion typically leverage direct N-body integrators \citep{lecaraarseth86,beaugeaarseth90,chambers01,kominamiida04,raymondetal09}. 
%Hybrid statistical algorithms also boast many of the same advantages of coagulation algorithms and N-body codes \citep{kenyonbromley06}, and they are best applied in the transition between different growth regimes.

Statistical or semi-analytical coagulation studies model the dynamics and collisions of planetesimals in a ``particle-in-a-box approximation'' \citep{greenbergetal78}. This method is based on the kinetic theory of gases.  It uses distribution functions to describe planetesimal orbits, and thus neglects the individual nature of the particles. This approach is routinely used to model the early stages of planet formation when the number of planetary objects is large ($>>10^4$). While these types of calculations give a statistical sense of the dynamics of a large population of gravitationally interacting objects, they also invoke a series of approximations which are only valid at local length scales in the \hbindex{protoplanetary disk} \citep{goldreichetal04}. The necessity of including non-gravitational effects and collisional evolution typically leads to approaches that are not self-consistent \citep{leinhardt08}.

For most applications, direct N-body numerical simulations tend to be more flexible and precise than coagulation approaches.  Until recently, N-body codes could not handle more than a few thousand self-interacting bodies for long integration times (e.g. $\sim10^8-10^9$ yr) without prohibitively long computational times.  However, recent advances in parallel computing have made calculations as many as $10^{6}$ particles computationally tractable \citep{grimmstadel14,menon15,lau23}.  There are several numerical N-body integration packages available to model planetary formation and dynamics such as Mercury \citep{chambers99}, Symba \citep{duncanetal98}, Rebound \citep{reintamayo15,reinspiegel15,tamayo20}, GENGA \citep{grimmstadel14,grimm22} and PKDGRAV \citep{stadel01}. Among them, Mercury and Symba are arguably the most widely used in the terrestrial planet formation literature.  These codes are built on symplectic algorithms which divide the problem's full Hamiltonian into a component describing the Keplerian motion, and a second Hamiltonian that carries all of the terms that arise from the mutual gravitational interactions between bodies in the system \citep{wisdomholman91}.  This approach is advantageous when applied to systems where most of the total mass is carried by a single body, and they  can facilitate long-term numerical integrations without the propagation and accumulation of errors \citep{sahatremaine94,hernandez17,rein19}. 
%While purely symplectic algorithms require a fixed integration step-size, close-encounters and collisions that occur frequently in planet formation simulations cannot be properly resolved without extremely small timesteps \citep{chambers99}. Thus, using a pure symplectic algorithm for planet formation applications with a timestep small enough to resolve planetary encounters would destroy its speed advantage. There are two main solutions to this problem. Symba divides up the full Hamiltonian and uses a different step-size for each part \citep{duncanetal98}. Mercury, GENGA and REBOUND propagate the orbits of bodies involved in close encounters with a different numerical integrator and a self-adaptive timestep \citep{chambers99}, while the remaining terms are solved symplectically.  Mercury and Symba are also routinely modified from their ``off the shelf'' forms to account for a range of phenomena such as nebular gas effects \citep[][others like GENGA include a gas disk perscription in the standard distribution]{mandell07,morishimaetal10,izidoro17}, collisional fragmentation \citep{chambers13,deienno19,cambioni21,childs22}, tidal migration \citep{bolmont15} and \hbindex{pebble accretion} \citep{kretkelevison14,bitsch19,lambrechts19,ogihara20,izidoro21_super_earth}.

Another way to simulate a system with a large number of particles is by combining direct N-body integration with super-particle approximation \citep{levisonetal12,morishimaetal15}.  In this approach, a large number of small particles (planetesimals) are represented by a small number of tracer particles. The tracers interact with the larger bodies though an N-body scheme. Tracer-tracer interactions (i.e., interactions among a large number of massive planetesimals represented by the tracer particles) are solved using statistical routines modelling stirring, \hbindex{dynamical friction} and collisional evolution. The LIPAD code \citep{levisonetal12} has been used, for example, to model the formation of terrestrial planets in the Solar System from a larger number of planetesimals \citep{walshlevison16,walsh19,deienno19} and also in  simulations that include \hbindex{pebble accretion} \citep[e.g.][]{levisonetal15pnas}. 

\section{Late stage accretion of terrestrial planets in the Solar System}

This section reviews models of the late stage accretion of terrestrial planets in our own Solar System. The first series of subsections discuss the constraints on these models, and the latter subsections presents different scenarios that can match these constraints.  The final text discusses strategies to distinguish or falsify these models.

\subsection{Solar System Constraints}

%Space observations, geological and cosmochemical analysis, computer simulations, etc., constrain  models of terrestrial planet formation. In this Section we discuss some of the key Solar System constraints to models of accretion of the terrestrial planets.

% Our intention is not to compile a very complete list of constraints but only provide a concise list. Any model satisfying the below mentioned constrains would be considered a viable scenario for the origins of the Solar System nowadays.
%Of course, some constraints are stronger or better understood  than others as for example those naturally observed. This is the case for example of the masses and orbits of the terrestrial planets. Other constrains are relatively weaker. For example, there is water on Earth but the exact amount of water on Earth's interior is still heavily debated. 

\subsubsection{Planetary masses, orbits and number of planets}

The total number of planets, their masses and orbits are typically viewed as the strongest constraints for formation models. While Mercury and Mars have moderately excited orbits ($e=$0.21 and $i=$7$^{\circ}$ for Mercury and $e=$0.09 and $i=2^{\circ}$ for Mars), those of Earth and Venus and quite circular.  Angular Momentum Deficit (\hbindex{AMD}) is commonly employed as a metric for quantifying the level of dynamical excitation of a planetary system \citep{laskar97,chambers01}. \hbindex{AMD} measures the fraction of a planetary system's angular momentum that is missing compared to a system where the planets have the same semi-major axes but circular and coplanar orbits.  \hbindex{AMD} can thus serve as a diagnostic of how well simulated terrestrial systems match the real inner planets' level of dynamical excitation, and is defined as:
 \begin{equation}
{\rm AMD}= \frac{{\sum_{j=1}^N}\Big[ m_j \sqrt{a_j} \> \Big( 1 - \cos i_j \> \sqrt{1-{e_j}^2}\> \Big)\Big]} 
{{\sum_{j=1}^N}\> m_j \sqrt{a_j}}.
\end{equation}
where $m_j$ and $a_j$ are the mass and semi-major axis of the $j$th planet, $N$ is the number of planets in the system, and ${\rm e_j}$ and ${\rm i_j}$ are the orbital eccentricity and inclination of each planet. The terrestrial planets' \hbindex{AMD} is 0.0018.  

Given the aforementioned excitation dichotomy in the inner Solar System (low excitation for Earth and Venus versus moderate excitation for Mercury and Mars), it can be potentially misleading to report the distribution of \hbindex{AMD}s that result from a suite of terrestrial planet formation simulations.  Indeed, a system containing an overly excited Earth analog and under-excited analogs of the other planets could potentially have the same \hbindex{AMD} as the Solar System.  For this reason, recent studies have simply used the final eccentricities and inclinations of Earth and Venus as metrics for success \citep{nesvorny21_tp,clement23,lykawka23}, as Mercury and Mars' eccentricities and inclinations are easier to reproduce, and thus tend to be less diagnostic \citep{brasseretal09,roig16,kaibchambers16}

Another useful metric is the Radial Mass Concentration (\hbindex{RMC})~\citep{chambers01}, a measure of a planetary's system's degree of radial concentration. Earth and Venus contain more than 90\% of the terrestrial planets' total mass in a narrow region between 0.7 and 1 AU. \hbindex{RMC} is defined as : 
\begin{equation} 
{\rm RMC} = {\rm Max}\left(\frac{{\sum_{j=1}^N}\> m_j}{{\sum_{j=1}^N}\> m_j\big[\log_{10}\left(a/a_j\right)\big]^2}\right). 
\end{equation}
Higher values indicate more concentrated systems. The inner Solar System \hbindex{RMC} is 89.9.  However, much like \hbindex{AMD}, \hbindex{RMC} is degenerate in the sense that a large number of small planets could have the same \hbindex{RMC} value as one with a small number of large planets.  Thus, a wide range of often complex simulation success criteria have been used throughout the literature.  Most of these schemes assign semi-major axis and mass limits to each planet, and consider simulations successful or marginally successful depending on how many inner planets are well reproduced \citep{clement18,lykawka19,nesvorny21_tp}.  In general, planets larger than 0.5-0.7 $M_{\oplus}$ are considered to be successful Earth and Venus analogs, 0.25-0.3 $M_{\oplus}$ is typically used as an upper limit for Mars' mass, and Mercury analogs smaller than $\sim$0.1-0.2 $M_{\oplus}$ are deemed satisfactory.

\subsubsection{The Asteroid belt}

%STOPPING HERE ON PROOF 16NOV 2PM

\begin{figure}
\centering
% \includegraphics[scale=.5]{real_ecc.pdf}
% \includegraphics[scale=.5]{real_inc.pdf}
% \caption{Orbital architecture of the inner Solar System. In both panels, planets are shown as triangles and asteroids as squares. The left-hand panel show a diagram semi-major axis versus orbital inclination. The right-hand panel shows orbital semi-major axis versus orbital inclination. Only asteroids larger than 50 Km are shown.}
\includegraphics[width=\textwidth]{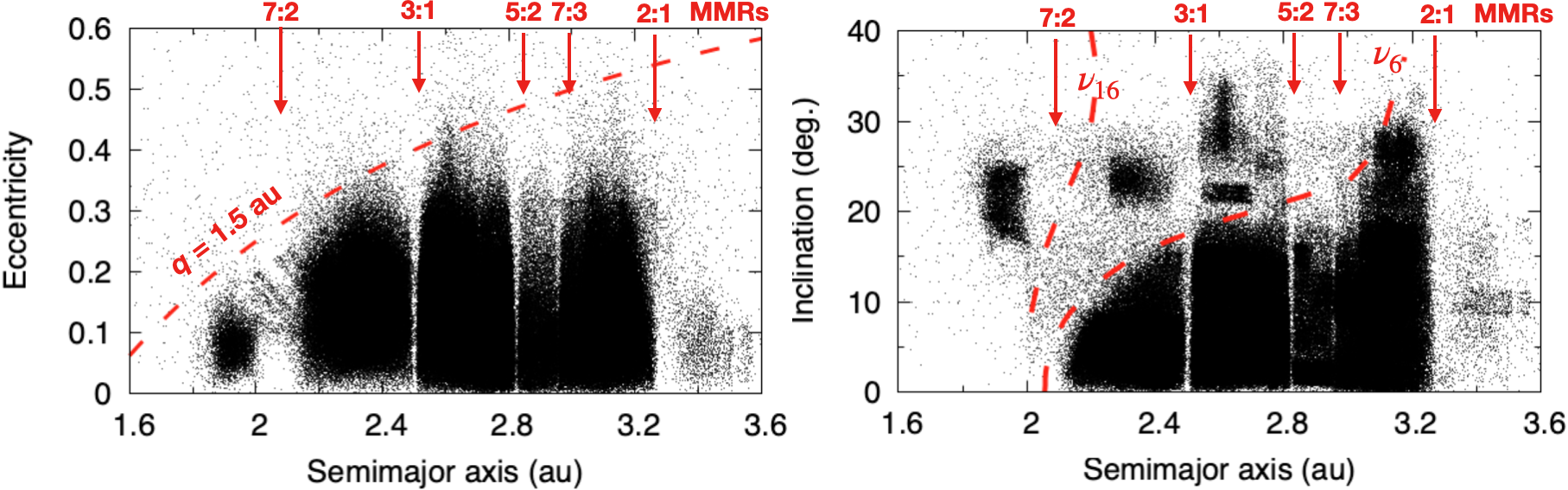}
\caption{Eccentricity (right) and Inclination (left) {\it versus} semi-major axis of objects with H $\le$ 17.75 (D $\approx$ 1 km assuming an average geometric albedo $\langle p_v\rangle$ = 0.14 taken from the MPC catalog).  The red arrows at the top of the plot indicate the position of the most prominent \hbindex{mean motion resonance}s (MMRs) with Jupiter, with the 2:1 MMR roughly delimiting the asteroid belt's outer boundary. The Curved dashed line in the left plot demarcates the perihelion distance where asteroids cross Mars' orbit (q = 1.5 au) that roughly limits the inner edge of the main belt eccentricity-wise. The approximate positions of the Saturnian \hbindex{secular resonance}s ($\nu_{16}$ and $\nu_{6}$) are denoted in the right panel with dashed red lines.}
\label{fig:3}       % Give a unique label
\end{figure}

Terrestrial and giant planets in our Solar System are physically separated by the asteroid belt. Unlike the reasonably circular and co-planar orbits of the planets, asteroids inhabit orbits that are quite dynamically excited. Asteroid eccentricities range from 0 to $\sim$0.4, and their orbital inclinations extend between 0$^{\circ}$ and $\sim$ 30$^{\circ}$ (essentially populating all non-planet-crossing orbits in the region: Figure \ref{fig:3}).  This dichotomy is expected to be a direct result of a giant planet instability that occurred some time after nebular dispersal \citep[][discussed in further detail later in this section]{roig15,clement19_ab}.

%The belt's inner edge is at around 1.8 au in semi-major axis space, and 1.5 au in perihelia space (Figure \ref{fig:3} left panel).
Several regions of the belt are traversed by powerful \hbindex{mean motion resonance}s (MMRs), along with two major \hbindex{secular resonance}s (SRs) that arise from perturbations from Saturn. The most prominent MMRs overlaying the asteroid belt are associated with Jupiter's  motion; occurring when the orbital period of an asteroid is an integer multiple of Jupiter's orbital period.  In contrast, SRs occur when an asteroid and planet's (Saturn, in the case of the main belt) precession frequencies are commensurable. As SRs overlap each of the dominant MMRs, asteroids in these regions are destabilized.  This process generates the so-called \hbindex{Kirkwood Gaps} in the belt's radial distribution (red arrows in Figure \ref{fig:3}).  Gaps in the belt's inclination distribution (dashed red lines in the right panel of figure \ref{fig:3}) are largely related to the migration of these resonances early in the solar system's history \citep{nagasawa05,walshmorbidelli11,clement20_nu6}.  In particular, the most important SRs in terms of past sculpting of the belt and current production of near-Earth asteroids are the $\nu_{6}$ (related to Saturn's eccentricity vector precession) and $\nu_{16}$ (driven by Saturn's nodal precession).  The outer limit of the asteroid belt is often associated with the 2:1 MMR at around 3.2 au.  Moreover, the distribution of objects about the major MMRs \citep{deiennoetal16} and SRs \citep{clement20_nu6} are strong diagnostic tools for constraining the past evolution of the giant planets. 

The asteroid belt is also quite devoid of mass when compared to the solar system's planetary regimes \citep[e.g.][]{petitetal01,petitetal02,morbidellietal15b}. The total mass of the four terrestrial planets is about 2$M_\oplus$.  In contrast, the main asteroid belt region only contains around $5\times 10^{-4}\mearth$ \citep{gradietedesco82,demeocarry13,demeocarry14}. Ceres is the most massive object in today's belt. Given the absence of large gaps in the belt's orbital structure that do not readily associate with MMRs or SRs \citep{raymondetal09,obrien11}, it is unlikely that the belt ever hosted objects larger that the Moon.
The origin of the asteroid belt's low mass is still a matter of intense debate \citep{morbidellietal09,walshetal11,deiennoetal16,raymondizidoro17a,clement19_ab}. Nevertheless, the belt's extreme low mass is still one of the strongest constraints for planetesimal, terrestrial planet, and giant planet formation theories (discussed in the subsequent section).

The asteroid belt is also chemically segregated \citep[e.g.][]{demeoetal15}.  The inner region is mostly populated by silicaceous asteroids (\hbindex{S-type}), while the outer region is dominated by carbonaceous asteroids (\hbindex{C-type}).  Thanks in part to the success of the \hbindex{Hayabusa} and \hbindex{OSIRIS-REX} sample return missions \citep{hayabusa1,osirisrex,hayabusa2}, there is broad agreement across multiple fields that \hbindex{S-type}s are associated with non carbonaceous chondrites (NC), and \hbindex{C-type}s associate with carbonaceous chondrites (CC). While a variety of other taxonomic asteroid classes exist, S- and \hbindex{C-type}s are by far the most abundant \citep{demeocarry14}. Due to their large inventories of carbon and hydrates, \hbindex{C-type} asteroids are quite dark. In contrast, \hbindex{S-type}s are moderately bright and mostly composed of rocks and iron \citep{gradietedesco82,demeocarry14}. S- and \hbindex{C-type} asteroids are expected to originate in orbits interior and exterior to that of Jupiter, respectively, and were most likely deposited in the asteroid belt region at later stages via scattering by growing terrestrial and giant planets \citep{raymondizidoro17a,raymondizidoro17b}.

\subsubsection{Water on Earth and other terrestrial planets}

The amount of water on Earth is uncertain \citep[e.g.][]{drakecampins06}. Estimates suggest that Earth's total water content is between $\sim$1.5 and $\sim$10-40 Earth oceans \citep{lecuyeretal98,marty12,halliday13}, where 1 Earth ocean is the total amount of water on Earth's surface (or $1.4\times 10^{24}g$; this includes all lakes, ice caps, glaciers and oceans).  A major fraction of this water is stranded in the Earth's mantle. Even more water may exist in Earth's core, but the true amount is much more difficult to constrain than that of the mantle \citep{nomuraetal14,badroetal14}. There is also evidence for water on Mercury \citep{lawrenceetal13,ekeetal16}, and the high \hbindex{D/H ratio} in Venus' atmosphere has been interpreted to strongly suggest that the planet once possessed a larger inventory of water that subsequently escaped to space \citep{donahueetal1982,kastingpollack83,grinspoon93}. Similarly, the high \hbindex{D/H ratio} of Mars' atmosphere, coupled with isotopic analyses of martian \hbindex{meteorites} also implies that some of its primordial water was lost to space \citep[e.g.][]{owenetal88,kurokawaetal14}. Geomorphological features on Mars indicate that the planet had ancient oceans and that a substantial amount of water is likely hidden below the surface \citep{bakeretal91}. All this evidence supports the idea that a significant amount of water was present in the inner Solar System during its formation.

Until recently, it was generally thought that asteroids in the inner region of the asteroid belt were drier than Earth. This implied that water needed to be `delivered' to an otherwise-dry planet; thus stimulating studies related to how planetesimals from beyond the asteroid belt (or even beyond Jupiter and Saturn's orbits) could have been transported to the terrestrial planet region \citep[e.g.][]{morbidellietal00,raymondetal04,raymondetal07,izidoroetal13,obrienetal14,raymondizidoro17a,morbidellietal12,obrienetal18,meech2020}. 

Isotopic ratios in \hbindex{meteorites} are a powerful tool that have been employed to better discriminate between water sources.  CC \hbindex{meteorites} are associated with \hbindex{C-type} asteroids in the belt and their hydrogen and nitrogen isotopic ratios -- D/H and ${\rm ^{15}N/ ^{14}N}$-- come close to matching those of Earth \citep{martyreika06,marty12}. The \hbindex{D/H ratio} of the solar nebula is generally inferred from Jupiter's atmosphere, and it is estimated to be about a factor of $\sim$5-10 lower than that of CCs.  Water with a \hbindex{D/H ratio} similar to the solar value  has been found in Earth's deep mantle \citep{hallisetal15}, but in order for Earth's water to have a primarily nebular origin one must invoke a mechanism to increase the \hbindex{D/H ratio} of Earth's water over the planet's history. In principle this could be achieved if the Earth had a massive primordial hydrogen-rich atmosphere that efficiently escaped to space over a billion year timescale due to the young Sun's very intense UV flux \citep{gendaikoma06,gendaikoma08}. However, the  solar ${\rm ^{15}N/ ^{14}N}$ ratio also does not match that of Earth \citep{marty12}. Comets present a wide range of \hbindex{D/H ratio}s, which vary from terrestrial-like to several times higher \citep{alexanderetal12}. Nevertheless, elemental abundances and mass balance calculations based on $^{36}$Ar suggest it is unlikely that comets contributed more than a few percent of Earth's water \citep{martyetal16}, but this same analysis also concluded that they probably contributed noble gases to Earth's atmosphere \citep{martyetal16,aviceetal17}. Therefore, until recently a consensus existed favoring CCs as the best candidates for delivering water to Earth \citep{alexanderetal12}. The much higher \hbindex{D/H ratio}s of Venus and Mars probably do not represent their primordial values and the origins of their water thus remains largely unconstrained. However, any process delivering water to Earth would invariably also deliver water to the other terrestrial planets \cite[e.g.][]{morbidellietal00,raymondizidoro17a}.

This classic paradigm of water delivery has recently begun to be reexamined in light of new isotopic measurements of \hbindex{meteorites} with improved sensitivity \citep[e.g.][]{izidoropiani2022}.  Recent isotopic analyses have shown that Enstatite chondrite (EC) \hbindex{meteorites} -- which are thought to have accreted near Earth's orbital distance -- contain far more water than previously estimated and actually boast water contents extremely similar to that of the modern Earth \citep{piani2020}.  The \hbindex{D/H ratio} of ECs comes about as close to matching Earth's as CCs \citep{piani2020}, making it difficult to disentangle the source of Earth's water.  When isotope ratios from other volatiles are taken into account -- notably nitrogen and zinc -- it has been argued that Earth's volatile budget can be explained as a 70-30 mix of ECs and CCs \citep{steller2022,savage2022,martins2023}.  If this was the case, given their higher concentration of volatiles, only a 5-10\% contribution of CC-derived material is required to meet the constraints from the Earth's total mass budget, with the rest of the planet's mass consistent with an NC source \citep{steller2022,savage2022,martins2023}.  

In this new paradigm, the majority of Earth's water would have been homegrown, and incorporated along with the NC building blocks that make up the bulk of our planet's mass.  Only a small fraction of Earth's bulk material originated in the outer Solar System, although it contributed a non-negligible portion of Earth's water and other volatiles. In contrast, Mars's volatiles follow a different trend, and are consistent with a much smaller contribution from CCs \citep{kleine2023}.  Future investigations into the volatile evolution of various meteorites and planetary bodies will be crucial for further pinning down the relative contributions of different reservoirs to the modern Earth's water content \citep[e.g.][]{petersonetal23}.

%If the Earth accreted mainly from rocky material exposed to the relatively high temperatures in the \hbindex{protoplanetary disk} than that material that accreted asteroids at larger distances it is reasonable to expect that the Earth  should be at least as reduced in  volatiles and water as  the innermost asteroids. Thus, given the larger amount of water on Earth it is believed that one or more mechanisms contributed delivering a major part of its water \citep[e.g.][]{morbidellietal00,raymondetal04,raymondetal07,izidoroetal13,obrienetal14,raymondizidoro17a}. 
%Interestingly, the Earth contains more water than would be expected from the radial water gradient across the Solar System \cite[see the recent review by][]{obrienetal18}. 

\subsubsection{Giant planet orbits and evolution}
\label{section:Nice_Model}

Numerical simulations and radiometric dating of materials from the Earth-Moon system demonstrate that last giant impact on Earth took place between $\sim$30 and $\sim$150 Myr after the formation of \hbindex{CAIs} \citep{yinetal02,jacobsen05,toubouletal07,allegreetal08,hallidayetal08,kleineetal09,rudge10,jacobsonetal14,fischer18,zube19}.  Mars, however, is probably much older than the Earth. Radiometric dating of Martian \hbindex{meteorites} using the Hafnium-Tungsten (Hf-W) isotope system indicate that Mars reached about half of its current mass during the first 2 Myr after CAI formation \citep{dauphaspourmand11}.  However, given the scarcity of Martian samples and the fact that dating methodologies are dependent on internal differentiation models, the planet's growth history is necessarily harder to pin down than that of the Earth.  Indeed, recent works have argued for a more protracted phase of accretion \citep{zhang21}.  Nevertheless, a majority of studies in the literature generally agree that Mars formed within just a few Myr after the appearance of \hbindex{CAIs} \citep{nimmokleine07,kruijer17_mars,costa20_mars}.  Therefore, the dichotomous nature of Earth and Mars' growth timescales presents a peculiar constraint for terrestrial planet formation models to match.  \hbindex{Meteorites} originating from Venus and Mercury have not been identified, thus making their ages unconstrained.  
%However, the absence of an internally generated dynamo and natural satellite at Venus have been interpreted as evidence that it did not experience any large impacts like the one that formed the Moon late in its accretion \citep{jacobson17}.

Even if the Moon-forming impact occurred around the earlier end of the window predicted by radiometric dating \citep[e.g. about 30 Myr after gas disk dissipation:][]{kleineetal09,rudge10,fischer18}, it is still widely accepted that the final phase of terrestrial accretion occurred long after the giant planets were fully formed  \citep{briceetal01,haischetal01}.  Given their large gaseous envelopes \citep{mizuno80,wetherill90,lissauer93,boss97,guillotetal04}, the giant planets are constrained to have formed prior to the dispersal of the gaseous \hbindex{protoplanetary disk} \citep{bodenheimerpollack86,pollacketal96,alibertetal05}. Therefore, virtually all models of terrestrial planet formation agree that giant planets play a critical role shaping the makeup of planets in the inner Solar System \citep[e.g.][]{wetherill78,wetherill86,chamberswetherill98,agnoretal99,morbidellietal00,chambers01,raymondetal06,obrienetal06,lykawakaito13,raymondetal14,izidoroetal14,
fischerciesla14,clement18}. 
Determining where and when to include giant planets in these models is challenging because they do not currently inhabit the same orbits that they were born with \citep{fernandezip84,malhotra93,hahnmalhotra99,massetsnellgrove01,tsiganisetal05,morbidellicrida07}.  
%While the general steps involved in how the giant planets attained their modern orbits are loosely agreed on, many of the details remain strongly debated.  For this reason, the precise nature of Jupiter and Saturn's early evolution are defining features of two of the three terrestrial planet formation models presented later in this section. 

Hydrodynamical simulations show that the giant planets' orbits probably migrated during the gas disk phase.  The most likely outcome of migration is a chain of \hbindex{mean motion resonance}s between successive planets \citep{massetsnellgrove01,morbidellicrida07,dangelomarzari12,pierensetal14}.  The current consensus model for the early dynamical evolution of the outer Solar System \citep[often reffered to as the \hbindex{Nice Model}:][]{gomesetal05,tsiganisetal05,morbidellietal05} argues that a number of dynamical aspects of the outer Solar System are well explained if the giant planets were transported from their initial resonant orbits to their current ones through an epoch of dynamical instability.  During the instability, the giant planets' orbits evolve rapidly \citep{nesvorny11}, and they can temporarily obtain eccentricities that are much larger than their current ones \citep{batyginetal12,clement21_instb2}.  These temporary periods of heightened excitation and rapid radial migration can strongly perturb the orbits of other planets and small bodies in the system. Among others, the Solar System qualities that seem to be consequences of this violent dynamical event include the capture of co-orbital asteroids \citep{morbidellietal05,nesvorny13} and irregular satellites \citep{nesvorny07,nesvorny14a} at all four giant planets, the orbital and resonant architecture's of Kuiper Belt \citep{levison08,nesvorny15a,nesvorny15b,nesvorny16,kaibsheppard16,nesvorny21_e_nep} and the precise orbital excitation of the giant planet's orbits \citep{morbidelli09_secular,nesvornymorbidelli12,deiennoetal17,clement21_instb,clement21_instb2}.  

As originally conceived \citep{gomesetal05,levisonetal11}, the \hbindex{Nice Model} was assumed to occur in conjunction with the late heavy bombardment \citep{tera74}; a perceived delayed spike in cratering events on the Moon evidenced by a preponderance of Apollo-sampled basins with ages of $\sim$3.9 Gyr.  For this reason, nearly all classic models of terrestrial planet formation assumed that the giant planets were likely in a low-eccentricity, resonant configuration during terrestrial accretion \citep{raymondetal06,obrienetal06,raymondetal09,izidoroetal14,izidoroetal15,izidoroetal16}.  However, the late instability model can be problematic in such a scenario because the giant planets' excited and chaotically evolving orbits strongly influence the terrestrial region during the instability.  This can lead to planet collisions and ejections.  Typically, Mercury or Mars do not survive the event \citep{kaibchambers16}.

Over the past five years, a wide range of observational, geophysical and dynamical constraints have been interpreted to evidence the instability having occurred relatively early in the Solar System's history \citep[within the first $\sim$100 Myr:][]{toliouetal16,deiennoetal17}.  While not an exhaustive list, these include an ancient binary trojan of Jupiter that would not have survived collisional evolution in the primordial trans-Neptunian region in the case of a late instability
\citep{nesvorny18}, the distinctive inventories of \hbindex{HSEs} (Highly Siderophile Elements) of the Earth \citep{beckeretal2006,bottkeetal2010} and Moon \citep{dayetal2007,dayandwalker2015} that suggest a significant quantity of material was delivered to the system \textit{after} the end of the Moon's \hbindex{magma ocean phase} but \textit{before} the end of the Earth's, \hbindex{cratering record}s in the inner Solar System \citep{brasser20,nesvorny23_craters}, collisional families in the asteroid belt that appear to be almost as old as the Solar System \citep{delbo17,delbo19} and certain properties of the Kuiper Belt's orbital distribution \citep{ribeiro20,nesvorny21_e_nep}.  Additionally, from a geochemical standpoint it is still unclear whether or not a spike in cratering on the Moon occurred 3.9 Gyr ago.  Updated basin ages leveraging $^{40}Ar/^{39}Ar$ dating \citep{norman06,liu12,grange13,merle14,mercer15,boehnke16} display a broader spread of dates.  Additionally, newer imagery and gravity data from contemporary missions have revealed a number of older basins underlying the younger ones that were presumably sampled by the Apollo missions \citep{spudis11,evans18}.  Given these new results, recent terrestrial planet formation models increasing attempt to address the fact that the instabity might have occured at some point within the first 100 Myr after the Solar System's birth \citep{clement18,clement19_frag,clement21_earlyinst,nesvorny21_tp,desouza21,clement23,lykawka23}

\subsection{Solar System Terrestrial Planet formation Models}
\label{section:TP_Formation_Models}

Simulations of late stage terrestrial accretion typically start with a population of already-formed planetesimals and Moon- to Mars-mass planetary embryos. This scenario is consistent with models of the runaway and \hbindex{oligarchic growth} regimes \citep{kokuboida96,kokuboida98,kokuboida00,chambers06,ormeletal10,ormeletal10b,carter15,morishima17,walsh19,clement20_embryo,woo21_embryo} as well as those considering \hbindex{pebble accretion} \citep{johansenetal14,moriartyfischer15,levisonetal15pnas,chambers16,johansenlambrechts17}.  This initial distribution of material loosely correlates with a starting epoch around $\sim$3 Myr after CAI formation \citep{raymondetal09}. Therefore, most of these simulations are initialized with fully formed giant planets and assume that the gaseous \hbindex{protoplanetary disk} has just dissipated \cite[e.g.][]{chamberswetherill98,agnoretal99}.

The most important ingredient in terrestrial accretion models is simply the amount of available mass. A zeroth-order estimate of the Solar System's starting mass comes from the ``Minimum mass solar nebula  model'' \citep[\hbindex{MMSN}:][]{weidenschilling77,hayashi81,desch07,crida09}. The original \hbindex{MMSN} model inflates the current radial mass distribution of Solar System planets to match the solar composition (adding ${\rm H}$ and ${\rm He}$; \cite{weidenschilling77,hayashi81}).  These \hbindex{MMSN}-derived models typically suggest that the primordial Solar System contained $\sim 5M_\oplus$ of solid material between the orbits of Mercury and Jupiter \citep{weidenschilling77}.

\begin{figure}
% For example, with the graphicx style use
\centering
\includegraphics[scale=.65]{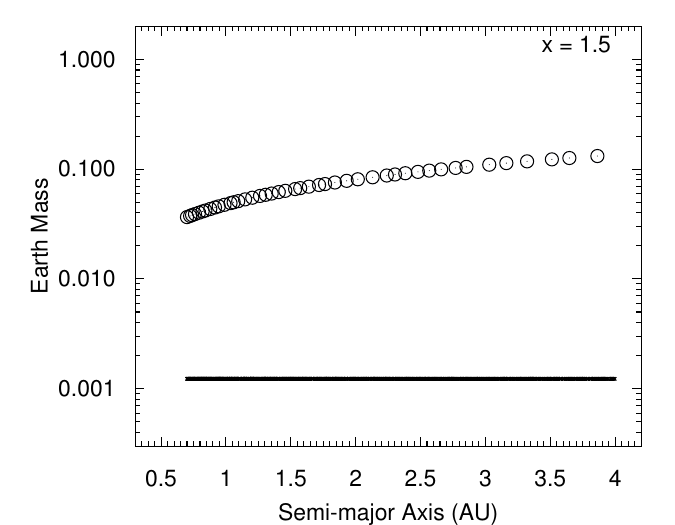}
\caption{Representative initial conditions for classic simulations of the late stages terrestrial planet formation using a power-law disk. In this case x=1.5 and ${\rm \Sigma_1 =8g/cm^2}$. The mutual separation of neighboring planetary embryos is randomly selected between 5 and 10 mutual \hbindex{Hill radii}. Planetesimals are shown with masses of $\sim 10^{-3}$ Earth masses. The total mass carried by about 40 embryos and 1000 planetesimals is about 4.5${\rm M_\oplus}$.}
\label{fig:4}       % Give a unique label
\end{figure}

Motivated by disk-formation simulations \citep[e.g.][]{bate18} as well as disk observations \citep[generally of the dust component;][]{andrewsetal10,willianscieza11}, the initial radial distribution of solids in simulations of late stage accretion typically follow power law profiles:
\begin{equation}
{\rm \Sigma(r)=
\Sigma_1 \left(\frac{r}{1 AU}\right)^{-x}\hspace{.3cm} g/cm^2 }.
\end{equation}
\label{eqn:powerlaw_disk}
${\rm \Sigma_1}$ is the surface density of solids at 1 AU. Initial planetary embryo masses are either derived from high-resolution simulations of planetesimal accretion and \hbindex{runaway growth} \citep{walsh19,clement20_embryo,woo21_embryo}, or by using the semi-analytically determined isolation mass that is proportional to  $r^{3(2-x)/2}\Delta^{3/2}$ \citep{kokuboida02,raymondetal05}.  Here, x is the power-law index and $\Delta$ represents the separation of adjacent planetary embryos in mutual \hbindex{Hill radii} \citep{kokuboida00}. A fraction of the disk total mass is typically distributed among equal-mass planetesimals \citep{raymondetal04,raymondetal06,obrienetal06}.  To improve compute times, planetesimals are often treated as ``semi-interacting'' particles; meaning that they can gravitationally perturb and collide with embryos but not with one another \citep{chambers01,raymondetal09}.  While the actual terrestrial disk's planetesimal inventory far exceeded the $\sim$10$^{3}$ used in most simulations of this type, studies varying the total number and individual masses of planetesimals find that they only result in lower order variations in final system outcomes \citep{jacobson14,clement20_embryo}. Figure \ref{fig:4} shows an example distribution of planetary embryos and planetesimals that follows a \hbindex{MMSN} disk profile.

%Models of late stage accretion of terrestrial planets has been build upon this idea. 

Multiple scenarios for the formation of the terrestrial planets have been proposed \citep{raymond20rev}.  The subsequent sections summarize the majority of these models, and highlights three models as potentially viable evolutionary paths for the inner Solar System.

\subsection{The Classic Scenario and the small-Mars problem}

The classic model assumes that giant planet formation and dynamical evolution can be completely disentangled from the process of terrestrial planet formation. Classic simulations simply impose a giant planet configuration (usually considering just Jupiter and Saturn) and a distribution of terrestrial building blocks. Early simulations in this mold succeeded in producing a few planets in stable and well separated orbits, delivering water to Earth analogs from the outer terrestrial disk and in explaining a significant degree of mass depletion of the asteroid belt~\citep{wetherill78,wetherill86,wetherill96,chamberswetherill98,agnoretal99,morbidellietal00,chambers01,raymondetal04}. Later, higher-resolution simulations were also able to reasonably match the terrestrial planets' \hbindex{AMD} and the timing of \hbindex{Earth's accretion}~\citep{raymondetal06,raymondetal09,obrienetal06,morishimaetal08,morishimaetal10}.

However, three major aspects of the systems formed in classic studies starkly contrast with the properties of the real terrestrial planets.  First, Mercury analogs typically grow to masses of $\sim$0.5-1.0 $M_{\oplus}$; in excess of an order of magnitude larger that the real planet.  While this problem is less severe if the disk's inner edge is depleted in mass prior to the onset of the giant impact phase \citep{chambers_cassen02,obrienetal06,lykawakaito17,clement21_merc_insitu}, the real Mercury mass lies at the extreme low end of the distribution of simulation-generated analog planet masses, and the majority of these worlds inhabit orbits that are too close to Venus.  In the extreme case where no massive particles inhabit the region interior to Venus' current orbit at time zero it is possible to form Mercury with a mass of $\sim$0.05 $M_{\oplus}$ \citep{hansen09,lykawka19,franco22}, however the proximity to Venus problem persists.  Given the relative dynamical isolation of Mercury's orbit, coupled with the fact that its large core mass fraction seems to evidence its mantle having been stripped via a high-energy impact \citep{benz86,benz07,asphaug14,chau18}, certain recent models have argued that the planets' peculiar mass and other properties are the result of dynamical interactions with the migrating giant planets \citep{raymondetal16,clement21_merc_lone_survivor,clement23} or terrestrial planet cores \citep{broz21,clement21_merc_outwardmig}.

In addition to their systematic shortcomings with regard to forming adequate Mercury analogs, classic terrestrial planet formation models such as those presented in \citet{chamberswetherill98} and \citet{raymondetal06} also tend to produce Mars analogs with masses much closer to that of Earth than that of the real planet (0.107 $M_{\oplus}$). While issues related to Mercury have received relatively sparse attention in the recent literature, this so-called small-Mars problem has sparked a prolific output of investigations and theorhetical models over the past two decades \citep{hansen09,walshetal11,lykawakaito13,jacobsonmorbidelli14,izidoroetal14,izidoroetal15,levisonetal15pnas,jacobsonetal14,clement18,clement19_frag,broz21,woo21_mars,izidoro22_natast,lykawka23}. In general, these models can be grouped into two classes: those that utilize dynamical mechanisms such as planet migration \citep{walshetal11} or dynamical instability \citep{clement18} to gravitationally perturb and remove material from the proto-Mars region and models that investigate the planetesimal formation process itself and conclude that few large bodies originated around Mars' modern orbit to begin with \citep{johansen21,morby21,izidoro22_natast}.  While all of these published models are capable of reproducing Mars' mass in a reasonable fraction of realizations, only a smaller subset have been scrutinized against a range of dynamical and cosmochemical constraints, and thus represent potentially viable formation models for the inner Solar System.  Perhaps the most critical of these constraints that is not matched in classic accretion models is the extremely rapid nature of \hbindex{Mars' formation} timescale relative to that of the Earth \citep{dauphaspourmand11,kruijer17_mars,costa20_mars}.  This leaves a relatively small window of opportunity for models using dynamical events to reshape the Mars region. 
This review focuses on three models that have been validated against a number of constraints. One of these models falls in to the class of models where the regions around Mars and the asteroid belt is initially \textbf{\textit{depleted}}, and the other two scenarios argue that the region was once full of material and subsequently \textbf{\textit{emptied}}.

The final persistent problem with classic terrestrial planet formation models is the propensity for $\sim$0.1-0.4 $M_{\oplus}$ planets to form in the asteroid belt \citep{chambers_wetherill01}.  Unlike the still mysterious case of \hbindex{Mercury's origin}, it is fairly obvious that issues pertaining to the mass budgets of Mars and the asteroid belt are inextricably linked \citep{izidoroetal15}. While it is certainly plausible that planetary-mass objects could have formed in the asteroid belt and been subsequently lost, the process would be problematic for the belt's orbital architecture.  Indeed, the fact that no significant gaps in the belt's $a/e$ and $a/i$ distributions that are not attributable to mean motion or \hbindex{secular resonance}s implies an upper limit on the mass of any object that could have ever formed in the asteroid belt of around a Lunar mass \citep{obrien11}. 

It is natural to seek out models where a single mechanism is responsible for either removing excessive material that would be responsible for forming an overly massive Mars along and unnecessary planets in the asteroid belt, or preventing the material from ever existing in the first place. The models discussed in the subsequent sections all adopt this approach of treating the Mars and asteroid belt mass deficiencies as facets of a common, fundamental problem with the classic model \citep{wetherill78,chamberswetherill98}. 

\subsection{The Grand Tack scenario}

Once planets grow large enough to exchange angular momentum with the gas disk \citep[somewhere between the masses of Mars and the Earth, depending on disk parameters:][]{papaloizou00,tanaka02,tanaka04} their semi-major axes can evolve inward or outward via Type-I or \hbindex{Type-II migration}.  In the former, lower-mass regime, a combination of torques that arise from the interactions between the star, disk and planet drive migration \citep{goldreich80,ward86,paardekoopermellema06,kleycrida08,paardekooperetal11,bentez15}.  In the latter case of Type-II evolution, planets massive enough to carve gaps in the nebular disk experience slower migration that is largely regulated by the radial gas flow within the disk \citep{durmann15,ida18}.  While the particular migration scheme experienced by a particular system is essentially impossible to determine as it is a complex function of a number of unconstrained properties of the primordial disk, many peculiar aspects of exoplanet demographics have been interpreted to strongly suggest that large-scale orbital migration has sculpted the majority of systems' dynamical architectures \citep[e.g.][]{ogihara09,lega15,ogihara15,izidoro17,bitsch19,lambrechts19,izidoro21_super_earth,izidoro22_radius_valley}.

Jupiter and Saturn's orbital migration can strongly perturb the orbits of objects in the inner solar system \citep{nagasawa05,thommes08,minton11}.  In the absence of clear constraints on the giant planet's actual birth location \citep[see][for recent papers that reach confliting results]{chambers21,dangelo21}, dynamical evolutionary models must consider each of four possibilities: (1) the gas giants formed close to their present locations, (2) they formed closer to the Sun and migrated out, (3) they formed further from the Sun and migrated in, and (4) migration was bidirectional.  While subsequent, post-disk evolution requires the giant planets roughly obtain semi-major axes close to their current ones around the time of gas dissipation \citep{tsiganisetal05,deiennoetal17,clement21_instb}, there are currently no strong constraints definitively ruling out any of these possibilities prior to disk dispersal \citep{massetsnellgrove01,pierensnelson08,pierenraymond11,morbyraymond16}. 

%I like all this but don't think we need this level of detail
%In the case of exclusively inward migration, the most significant consequences in the asteroid belt and terrestrial reservoir of planetesimals is icy planetesimal implantation \citep{raymondizidoro17a}.  \citep{deienno22} found that, even given the most extreme assumptions for initial asteroidal mass and subsequent depletion, if Jupiter had migrated inward from further than 15 au it would have implanted an excessive amount of mass into the belt that could not be removed by the present day.  Presumably, less-extreme inward migration schemes would still be allowable, and consistent with other solar system constraints.  The possibility of outward migration has received attention in the literature as well.  Indeed, certain giant planet formation models suggest that Jupiter-mass planets can form at very low radial distances via pebble trapping \citep{chatterjee14,boley14}, and large scale outward migration is favored for a range of disk parameters in hydrodynamical simulations \citep[e.g.][]{paardekooperetal11,bitsch15}.  If this were the case, depending on the timing and speed of the migration \citep{raymondetal16,clement21_merc_outwardmig} it would have significantly sculpted the inner disk and asteroid belt region.

The \hbindex{Grand Tack} invokes bidirectional migration to remove mass from the proto-Mars and asteroid belt regions \citep{walshetal11,walshetal12, raymondmorbidelli14,jacobsonetal14,brasseretal16,deiennoetal16,walshlevison16,allibert23,ogihara23}.  In essence, the model posits that Jupiter and Saturn followed a very specific inward-outward migration scheme that consequently sculpts the mass and compositional distribution of objects in the asteroid belt, while also truncating the terrestrial disk into a narrow annulus with an outer edge around $\sim$1.0 au. In successful simulations of the latter stages of giant impacts in the inner solar system, Mars is typically scattered out of the annulus and stranded on its modern orbit \citep{agnoretal99,hansen09}. This formation avenue for Mars also serves to limit the probability of large, late giant impacts on the planet; in good agreement with Hf-W constraints on its formation timescale \citep{dauphaspourmand11}.

\begin{figure}
% For example, with the graphicx style use
\hspace{-1.5cm}
\includegraphics[scale=1.5]{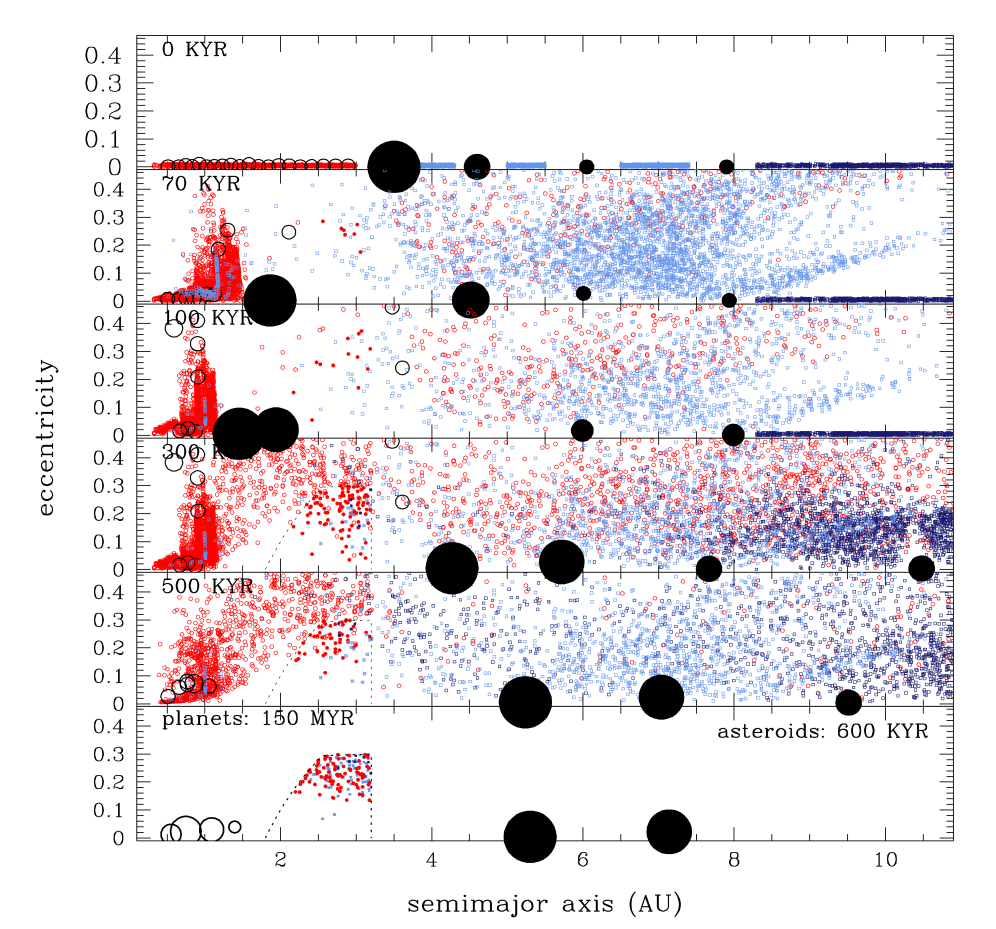}
\caption{Snapshots showing the dynamical evolution of the Solar System in the \hbindex{Grand Tack} model. The four gas giants are represented by the black filled circles. Jupiter starts fully formed while the other giants planets grow. Terrestrial planetary embryos are represented by open circles. Water-rich and water-poor planetesimals/asteroids are shown by blue and red small dots, respectively. During the inward migration phase Jupiter shepherds planetesimals and planetary embryos, thus creating a confined disk around 1 AU. Saturn encounters Jupiter and both planets start to migrate outwards at about 100 kyr. During the outward migration phase the giant planets scatter planetesimals inward and repopulate the previously depleted belt with a mix of asteroids originated from different regions. Figure reproduced from \cite{walshetal11}.} 
\label{fig:grand_tack}       % Give a unique label
\end{figure}

Figure \ref{fig:grand_tack} demonstrates how the \hbindex{Grand Tack} model provides a compelling explanation for the belt's diminutive mass, dynamically excited state, and observed radial mixing of S- and \hbindex{C-type}s \citep{walshetal12,deiennoetal16}. When Jupiter first encounters the inner planetesimal disk during its initial inward migration phases, it immediately disperses priomordial asteroid belt planetesimals (red points in figure \ref{fig:grand_tack}) throughout the solar system. Some objects are shepherded inward via smoothly migrating interior resonances with Jupiter, and others are scattered onto more distant orbits. Thus, after Saturn's inward migration is complete, the present day asteroid belt region is largely empty, and the outer solar system possesses a mixture of material originally derived from the inner and outer solar system (S- and \hbindex{C-type} objects). When the two gas giants reverse the direction of their migration, or ''tack,'' they encounter this sea of scattered planetesimals. The majority of these objects are further scattered onto more distant and more eccentric orbits, however a small fraction survive on stable orbits in the modern asteroid belt and some deliver water to the growing terrestrial planets through impacts \citep{obrienetal14,obrienetal18}.

\citet{deiennoetal16} used high-resolution numerical simulations to investigate the consequences of the combined effects of Jupiter's migration through the terrestrial disk, the outer solar system's subsequent epoch of dynamical instability \citep{tsiganisetal05,nesvornymorbidelli12}, and the ensuing $\sim$4.0 Gyr of evolution on the asteroid belt's eccentricity and inclination distributions.  In general, the distribution of modeled asteroid eccentricities provided a reasonable match to the real one.  However the inclination distribution of the inner main belt was overly excited.  A possible remedy for this deficiency might be the removal of excessive high-inclination objects during Jupiter and Saturn's post-instability migration as a result to resonance cycling caused by their conspicuous proximity to the 5:2 comensurability \citep{clement20_nu6}.

In addition to its success when measured against constraints related to Mars and the asteroid belt, internal structure models of Earth's growth and differentiation \citep{rubieetal15} have demonstrated that the model can also match Earth's internal composition and the mantle's inventory of water in broad strokes.  These studies \citep[see also][for a more recent investigation]{ogihara23} also predict a substantial fraction of water being incorporated into the core of the growing Earth; and thus provide a potential explanation for the apparent under-density of its core \citep{birch52}.

\subsection{The Early Instability scenario}

Jupiter need not physically traverse the terrestrial region to deplete planetesimals in the asteroid belt and proto-Mars regions.  Indeed, simply altering a classic terrestrial planet formation simulation \citep{chamberswetherill98} by placing Jupiter and Saturn on their modern orbits with eccentricities equal to double their current values regularly yield Mars analogs of the appropriate mass \citep{raymondetal09}.  Multiple studies \citep{nagasawa05,bromleykenyon17,woo21_embryo} have also demonstrated that, if the giant planets attain non-zero eccentricities during the gas disk phase \citep{pierensetal14,clement21_instb}, the locations of their \hbindex{secular resonance} sweep across the asteroid belt and inner Solar System during the disk's photoevaporation phase; thereby depleting the region and limiting the mass of Mars and the asteroid belt.  Moreover, \citet{lykawakaito13} found that a small Mars can be formed if Jupiter and Saturn migrate smoothly with elevated eccentritites from their formation configuration \citep[presumably the 3:2 \hbindex{mean motion resonance}:][]{morbidellicrida07,nesvornymorbidelli12} to their modern orientation with respect to one another.  These three results demonstrate the extreme sensitivity of planetesimals in the terrestrial region to changes in the giant planets' eccentricities and spacing.

The \hbindex{Early Instability} model attempts to circumvent a late instability's harmful effects on the fully formed terrestrial planets \citep{brasseretal09,agnorlin12,brasser13,kaibchambers16} by taking advantage of a wealth of recent constraints that convincingly pin the instability's occurrence down to the first 100 or so Myr after the Solar System's birth (discussed in the previous section).  \citet{clement18} investigated a range of possible timings within this window and found that, if the event occurs within the first 1-10 Myr after gas dispersal, resonant perturbations from the excited giant planets conspire to remove material from the region around Mars' orbit and the asteroid belt.  In addition to providing a resolution to the small Mars problem, this particular timing also fits in well with meteoritic constraints on \hbindex{Mars' formation} timescale \citep{dauphaspourmand11,kruijer17_mars}.  If the instability occurs too early, while the terrestrial region is still abundantly populated with small planetesimals, the disk that was truncated by the giant planets' excited orbits has a tendency to re-spread and produce under-mass Earth and Venus analgos and overly massive Mars analogs \citep{clement18}.  Contrarily, if the instability occurs too late, Mars has already grown beyond its modern mass, and the final planetary systems are similar to those formed in classic terrestrial planet formation models \citep{nesvorny21_tp}.  Thus, as far as the terrestrial planets are concerned, late instabilities (t$>$10 Myr) would only be viable if the regions around Mars and asteroid belt were already at least partially depleted in mass prior to the instability's onset \citep{nesvorny21_tp,lykawka23}.  An instability at t$\simeq$50-100 Myr would also occur around the same time as the Moon-forming impact \citep{kleineetal09}, and could potentially trigger the event \citep{desouza21}.

An early giant planet instability is a rather effective mechanism for removing a substantial amount of mass from a primordially massive asteroid belt \citep{deienno18}.  \citet{clement19_ab} noted that depletion in the belt scales as a function of Jupiter's eccentricity excitation.  When Jupiter's eccentricity is excited all the way to its modern value \citep[a constraint that itself is challenging to match in instability models, see][for a more detailed explanation]{nesvornymorbidelli12,clement21_instb}, the belt can be depleted by 2-3 orders of magnitudes.  However, this depletion factor is higher than what has been found in more recent studies ($\sim$90$\%$ depletion) using controlled instability evolutions \citep{deienno18,nesvorny21_tp}.  If the primordial terrestrial disk indeed possessed a smooth surface density profile \citep[as considered in][]{clement18}, it would have possessed approximately 10,000 times more mass initially than it does today.  Post-instability chaotic evolution and resonant interactions over $\sim$4 Gyr is only expected to deplete the belt by $\sim$50$\%$ \citep{minton10}.  This leaves three options for explaining the belt's present mass of $\sim$5.0x10$^{-4}$ $M_{\oplus}$, all of which are potentially compatible with the \hbindex{Early Instability} as an explanation for Mars' mass.  First, the asteroid belt could have been born massive \citep[consistent with a uniform surface density profile in the inner Solar System and the \hbindex{MMSN}:][]{weidenschilling77}.  This would require a rather strong instability as proposed in \citet{clement19_ab} to deplete the belt by around three orders of magnitude, however such a massive primordial belt is likely incompatible with constraints from the belt's SFD \citep{deiennoetal2024}.  Second, the asteroid belt could have been partially depleted prior to the instability, perhaps by a \hbindex{Grand Tack}-like migration of the giant planets \citep[not neccessarily as radially extensive as proposed in][]{walshetal11}.  A more mild instability like the one modeled in \citet{deienno18} could have then subsequently depleted the remainder of excessive mass in the region.  Finally, the asteroid belt could have been born with very little mass \citep{raymondizidoro17b}, and the instability was not responsible for its depletion at all \citep[note that these last two options have not been explicitly tested, however more simplified models presented in][suggest that they are reasonable]{clement19_frag,nesvorny21_tp}.

The dynamical excitation of the giant planets during the instability provides a mechanism that significantly mixes material between different radial bins \citep{clement19_ab} in the asteroid belt \citep{clement19_ab}, and simultaneously scatters distant, icy planetesimals inward onto trajectories that allow for water delivery to Earth \citep{clement18}.  Therefore, provided an alternative mechanism for implanting \hbindex{C-type} objects \citep[for instance, through aerodynamical destabilization during Jupiter and Saturn's growth phase][]{raymondizidoro17a}, the model is fully consistent with the asteroid belt's modern radial distribution of taxonomic classes.

%If Morby's science paper or DPS abstract becomes citable we will have to address it here
Recent work on the \hbindex{Early Instability} model has demonstrated its compatibility with Earth's core and mantle compositions \citep{gu23}, as well potentially viable formation avenues for Mercury \citep{clement23}.  In particular, the excited giant planets' orbits' dynamical perturbations in the inner Solar System facilitate high-velocity collisions that are capable of stripping Mercury's primordial mantle and reproducing the planets' large inferred core mass fraction \citep{hauck13}.

\begin{figure}
\centering
\includegraphics[scale=0.5]{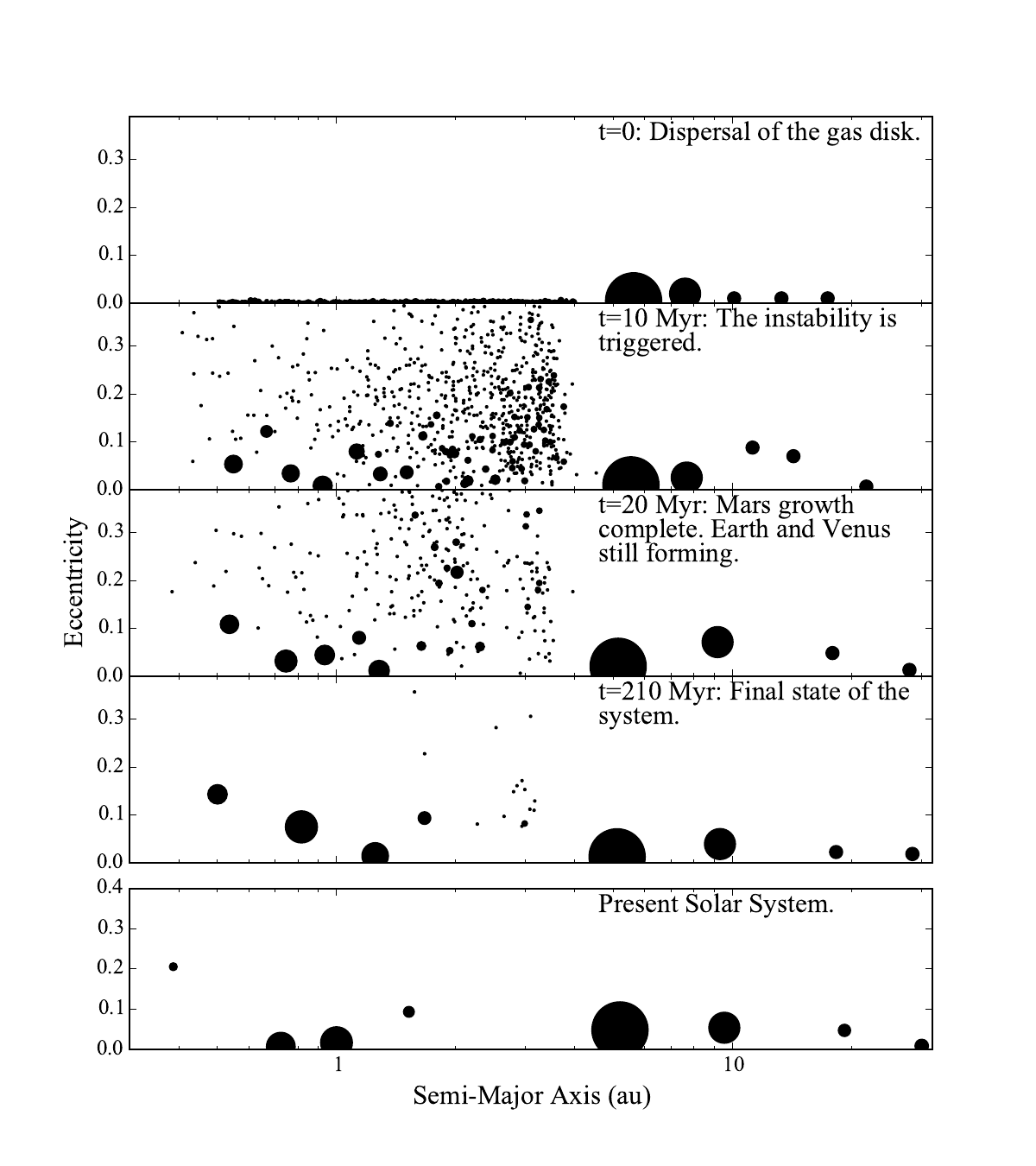}
\caption{Example successful evolution of the \hbindex{Early Instability} model \citep[reproduced from][]{clement18}.  The second panel shows the onset of the giant planet instability around $\sim$10 Myr after the dispersal of the gas disk \citep[note that more recent works have experiemented with alternative timings with mixed results]{clement19_frag,clement21_earlyinst,nesvorny21_tp}.  Perturbations from Jupiter and Saturn's excited orbits destabilize and remove material from the proto-Mars and asteroid belt regions, excite the orbits of asteroids that do survive \citep{deienno18,clement19_ab}, while leaving Earth and Venus' formation largely undisturbed (panels 3 and 4).}
\label{fig:early_instability}
\end{figure}

\subsection{Planet Formation from Rings and the primordially empty/low-mass asteroid belt}

While the first two models discussed here suggest that the asteroid belt was born massive and \textit{emptied} via dynamical perturbations from, or the physical presence of Jupiter and Saturn, it has also been proposed that the belt did not contain a substantial population of planetesials to begin with.  The simplest explanation for how this might have happened is by Jupiter's core blocking the inward pebble flux and thus starving much of the inner solar system of the raw materials needed to form large bodies ~\citep{lambrechts14,morby15}.  Indeed, recent high-resolution \hbindex{ALMA} observations of ring-like features within \hbindex{protoplanetary disk}s \citep{huang18} seem to indicate that the radial structures of actual young planet-forming regions are anything but smooth \citep[e.g.][]{hayashi81}.  Additional evidence for large-scale structure early in the solar system's history comes from the disparate compositions of NC and CC \hbindex{meteorites} (the so-called \hbindex{NC/CC dichotomy}) when measured using a number of different isotope systems \citep{budde16}.  These distinctive inferred formation ages and compositions have been interpreted to imply that two distinctive regions of planetesimals were physically separated early in the solar system's history \citep{kruijeretal17}.  Several explanations for this separation of reservoirs have been proposed, including the rapid formation of Jupiter's core \citep{kruijeretal17} and a primordial pressure maximum in the vicinity of Jupiter's modern orbit caused by an \hbindex{ALMA} disk-like ring feature \citep{brasser20_NC_CC}.

Returning to the aforementioned issues with terrestrial planet formation models, it is noteworthy that the terrestrial planet's orbits are extremely well reproduced when the initial distribution of planetesimals and embryos is confined within a narrow annulus between $\sim$0.7-1.0 au \citep{agnoretal99,hansen09,levisonetal15pnas,izidoroetal15,kaibcowan15,raymondizidoro17b,lykawka19}.  Until the observation of ring-like features in \hbindex{protoplanetary disk}s, the lack of an \textit{explanation} for the origin of a such a narrow radial feature was the biggest weakness of these so-called ``annulus'' models.  Given the plethora of \hbindex{ALMA} observations of ring-like structures, in tandem with the inferred \hbindex{NC/CC dichotomy} in the solar system, the challenge for contemporary modelers has shifted to finding an explanation for the formation of rings in very specific locations of the solar nebula.

Planetesimal formation is well known to be highly sensitive to the local disk properties \citep{simonetal16,yang17,yang21}.  Indeed, dust coagulation and \hbindex{pebble accretion} models have demonstrated that, given the correct selection of disk parameters, large concentrations of planetesimals can form around 1 au, but not elsewhere in the terrestrial region \citep{moriartyfischer15,drazkowska16}.  \citet{izidoroetal15} systematically studied the formation of terrestrial planets in disks with different radial mass distributions \citep[i.e. in shallow and steep surface density profiles, see also:][]{raymondetal05,kokuboetal06}.  This study identified an important trade-off between Mars' mass and the asteroid belt's level of excitation: shallow disks produce overly massive Mars analogs and properly excited asteroid belts while steeper disks typically yield good Mars analogs and under-excited asteroids.  The latter is a result of inefficient gravitational self-stirring in models of a severely depleted primordial belt, however this issue is resolved if the giant planet instability occurs earlier within the process of terrestrial planet formation \citep[][although not necessarily as early or within as specific a window as required in the \hbindex{Early Instability} model]{izidoroetal16,deienno18}.

Subsequent work on this scenario demonstrated that it is entirely possible no modern asteroids actually formed in the asteroid belt. Terrestrial planet formation simulations that are initialized with a narrow annulus of embryos and planetesimals naturally implant a small fraction of planetesimals from the terrestrial region into the asteroid belt \citep[in excess of the current total mass of \hbindex{S-type}s:][]{raymondizidoro17b,woo22}. During the chaotic sequence of giant impacts that unfolds during the terrestrial accretion process planetesimals are scattered onto high-eccentricity, belt-crossing orbits. A fraction of these are subsequently scattered by rogue embryos onto lower-eccentricity orbits beyond 2 AU, preferentially in the inner main belt. Thus, if the number of primordial asteroids that form in the belt is less than the present day region's constituency of \hbindex{S-type}s, it would imply that the majority of modern \hbindex{S-type}s are ``refugees'' from the a$<$2.0 au region of the disk.

\begin{figure}
\includegraphics[scale=.55]{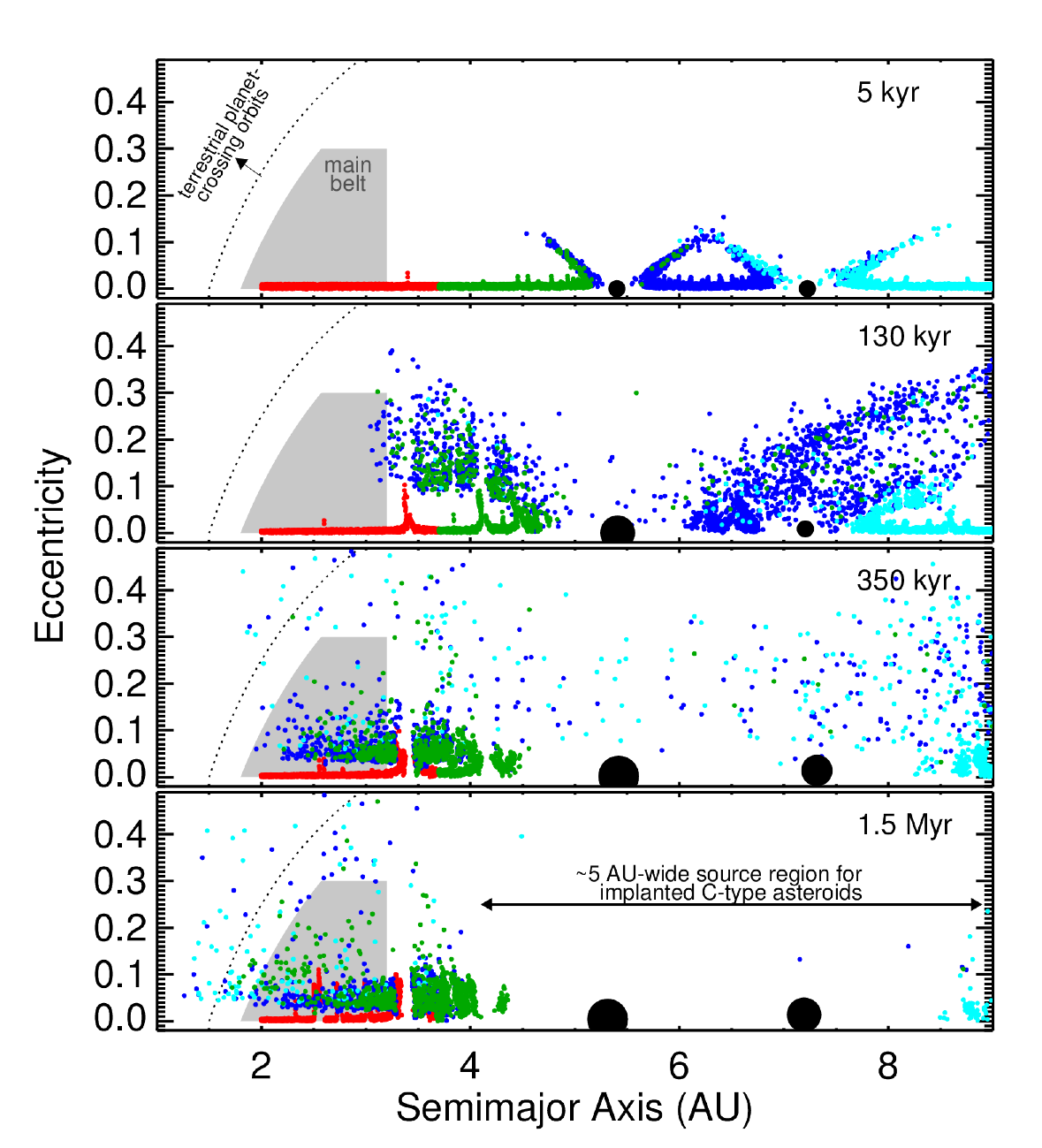}
\caption{Implantation of asteroids in the belt during the epoch of rapid gas accretion onto Jupiter and Saturn \citep[reproduced from][]{raymondizidoro17a}. The gas giants are represented by the growing filled circles. Planetesimals are color-coded to reflect their original locations.  The gray shaded region delimits the asteroid belt. In the depicted simulation, Jupiter grows linearly from a 3${\rm M_\oplus}$ core to its current mass between 100 and 200 ky, while Saturn's runaway accretion takes place between 300 to 400 kyr.}
\label{fig:implantation}
\end{figure}

A different process is likely responsible for implanting \hbindex{C-type} asteroids native to the outer solar system into the belt.  During the phase of giant planet growth where Jupiter and Saturn rapidly accrete gas directly from the nebula, the orbits of nearby planetesimals are perturbed and gravitationally scattered onto eccentric orbits. Given the dissipative nature of \hbindex{gas drag}~\citep{adachietal76}, the eccentricity of inwardly-scattered planetesimals are subsequently damped; leading to their capture onto stable orbits in the outer main belt (see Figure \ref{fig:implantation}). A fraction of these scattered bodies are not implanted into the belt region and instead acquire eccentricities that are large enough to allow their trajectories to cross those of the growing terrestrial planets and deliver water \citep{raymondizidoro17a}. In contrast with other models (e.g., the \hbindex{Grand Tack}), this mechanism is an {\it unavoidable} consequence of giant planet formation that would operate in any proposed terrestrial planet formation scenario.

Originally referred to as the ``low-mass asteroid belt'' model \citep{raymondizidoro17b}, a number of recent studies have designed extremely detailed models that investigate the combined effects of disk chemistry, gas dynamics, \hbindex{pebble accretion} and post-disk dynamics, and attempt to comprehensively link the earliest formation of ring-like structures with the solar system's modern architecture.  \citet{lichtenberg21} developed a one-dimensional disk model and argued that, under a reasonable set of assumptions, the migration of the water-ice \hbindex{snowline} can proceed in a manner such that planetesimal formation occurs in two distinctive bursts, at disparate locations and times.  In successful simulations, the inner region contains roughly an Earth-mass worth of planetesimals, and the total mass of large bodies in the outer reservoir is around that of Jupiter.  However, certain distinctive compositional properties of NC and CC \hbindex{meteorites} \citep[for example, their dissimilar isotopic ratios, oxidation states and Fe/Ni ratios, e.g.:][]{spitzer21} are challenging to reconcile in a model where their parent bodies form under similar conditions (i.e.: at the snow-line) but at different locations and times.

\begin{figure}
    \centering
    \includegraphics[width=.65\textwidth]{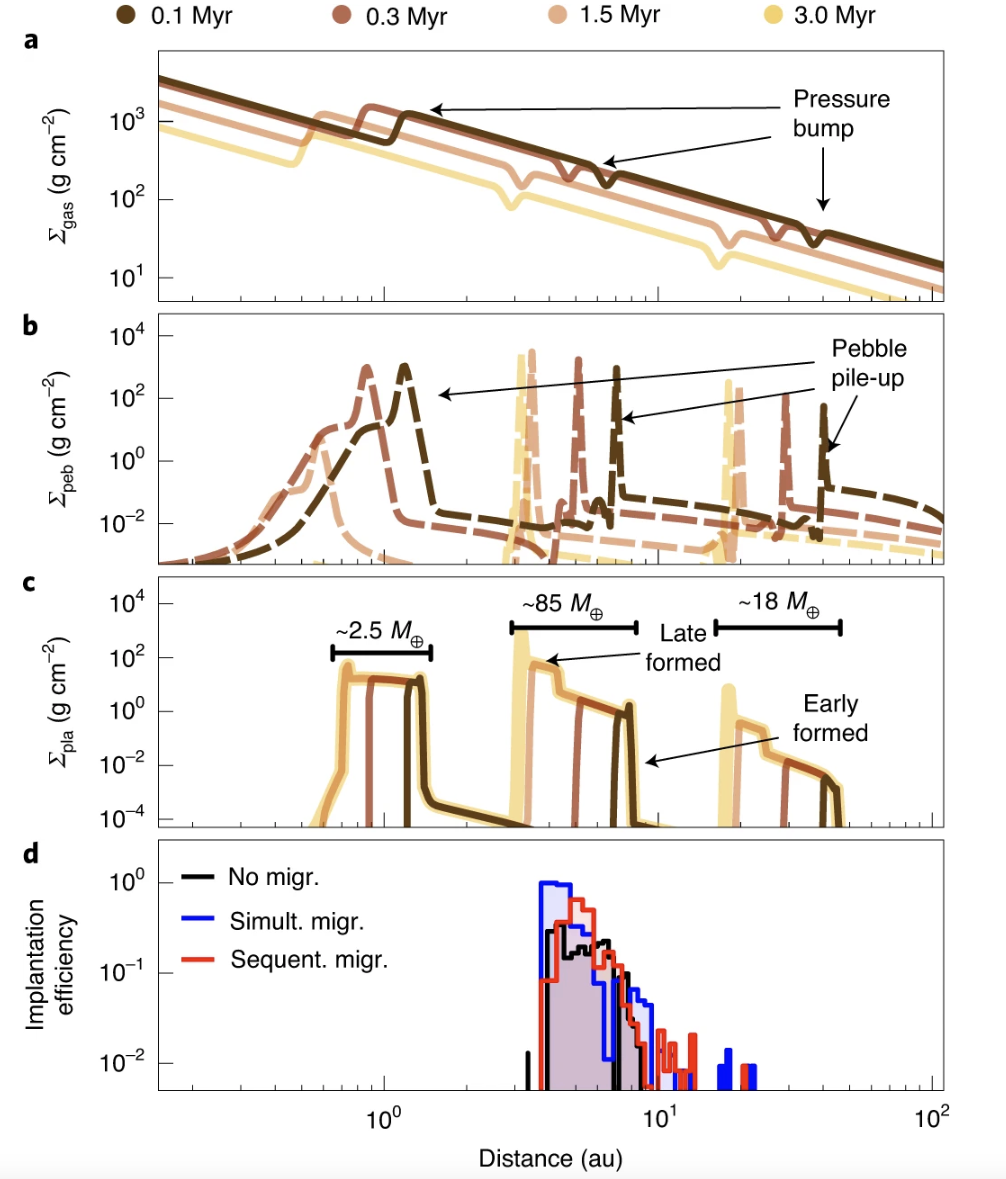}
    \caption{Example evolution where snow-line driven ring-like structures in the solar nebula sculpt the gas and solid radial surface density profiles into concentrated segments that match the present-day system's peculiar mass distribution \citep[reproduced from][]{izidoro22_natast}.  The top panel shows the assumed disk gas surface density profile at various times in the simulation (different shades of brown and yellow lines corresponding to 0.1, 0.3, 1.5 and 3.0 Myr).  The three primordial pressure bumps are linked to the locations of the Silicate sublimation (T $\simeq$ 1,400 K), the water snow-line (T $\simeq$ 170 K) and the CO snow-line (T $\simeq$ 30 K).  The features' affect on the solid pebble population is depicted in the second panel, and is ultimately responsible for triggering planetesimal formation in highly localized, ring-like regions.}
    \label{fig:rings_model}
\end{figure}

\citet{morby21} used a similar numerical approach as \citet{lichtenberg21} and found that, if turbulent diffusion is low, it is possible to trigger planetesimal formation within the first $\sim$500 kyr of evolution via dust pile-ups at both the \hbindex{snowline} (around $\sim$ 5 au) and the silicate sublimation line (closer to $\sim$ 1 au).  Most recently, \citet{izidoro22_natast} used a one-dimensional disk model accounting for dust coagulation, fragmentation, and turbulent mixing to show that, assuming the solar nebula contained a set of minor, primordial pressure bumps (top panel of figure \ref{fig:rings_model}) at the locations of the silicate sublimation line, water snow-line and CO snow-line, planetesimal formation proceeds via dust pile-ups in highly localized regions (second panel of figure \ref{fig:rings_model}).  Presumably, these pressure bumps would be the result of disk opacity  \citep{muller21,charnoz21} or zonal flows \citep{pinilla21}.  With only these assumptions, the ``ring'' model is able to generate the narrow annulus of terrestrial disk material \citep{hansen09} needed to reproduce the masses and formation histories of the terrestrial planets \citep{raymond20rev}, an appropriate quantity of planetesimals in the outer disk to form the giant planets and the Kuiper belt (third panel of figure \ref{fig:rings_model}), and the correct total masses and compositional dichotomy of S- and \hbindex{C-type}s in the asteroid belt \citep[bottom panel of figure \ref{fig:rings_model}:][]{demeocarry13}.  \citet{izidoro22_natast} also performed a series of N-body, late-stage accretion simulations to demonstrate the model's compatibility with multiple other inner solar system qualities such as the dissimilar isotopic compositions of Earth \citep[perdominantly EC:][]{dauphas17} and Mars \citep[mostly Ordinary Chondrites:][]{tang14}.  While additional work will be required to verify the model's compatibility with dynamical constraints from the asteroid belt's eccentricity and inclination distributions, it is reasonable to assume that it will be largely consistent with other studies that find the subsequent \hbindex{Nice Model} instability primarily sculpts these properties \citep{roig15,deiennoetal16,clement20_nu6}.

 \subsection{Models for the formation of Mercury}

%%%MSC LEAVIVNG OFF HERE FOR NOW NOV16 @ 9PM.  Only a bit more to go!

 As discussed in the previous section on solar system constraints, a successful model for \hbindex{Mercury's origin} must simultaneously explain why its mass is only 5-6$\%$ that of Earth and Venus, why its orbit is dynamically detached from those of the other terrestrial planets, and the reason for formation of its anomalously large core mass fraction \citep[CMF:][]{hauck13}.  Of these three issues, the latter has arguably received the greatest attention in the recent literature.  Specifically, there are two possible explanations for Mercury's CMF of $\sim$0.7-0.8.  In the first class of models \citep[``chaotic:''][]{benz88,ebelandstewart17}, proto-Mercury is hypothesized to be the victim of an extremely violent collision between two objects initially possessing Earth-like CMFs of $\sim$0.3.  As a result of the collision, Mercury's primordial mantle is stripped and the final planet is comprised mostly of the original object's core material.  In the second paradigm \citep[``orderly:''][]{morgan80,ebelandstewart17}, the planetesimals in the proto-Mercury region already posses boosted iron contents prior to the onset of the giant impact phase.  While both scenarios can occasionally replicate other aspects of Mercury (specifically its size and orbit), in contrast to the aforementioned models that explain Mars' low mass and rapid formation timescale, no model currently exists that can consistently match constraints for Mercury's CMF, mass and orbit, as well as other broad dynamical constraints for the rest of the inner Solar System.

\subsubsection{Chaotic Models}

Similar to studies of the Moon-forming impact \citep[e.g.][]{cameron85,benz86,canup04}, investigations of the giant impact hypothesis for \hbindex{Mercury's origin} typically leverage hydrodynamical models with the aim of uncovering impact geometries (initial masses and velocity vectors of the target and projectile) that can reproduce Mercury's current mass and CMF.  In addition to the planet's internal structure, these studies are motivated by the fact that Mercury's crust and mantle are highly reduced and depleted in volatiles \citep[however, the \hbindex{MESSENGER} mission revealed that Mercury's inventory of less volatile lithophile elements like Si, Ca, Al, and Mg is similar to that of the other terrestrial planets]{weider15,ebelandstewart17}.  Within this class of models, two general possibilities exist: proto-Mercury as the target being struck by a smaller projectile \citep{benz07} or proto-Mercury being the projectile and colliding with a larger proto-planet \citep{asphaug14}.  In the latter scenario, the obvious targets would be proto-Earth or proto-Venus.  These possibilities therefore require a mechanism for dynamically detaching Mercury's orbit from those of the other terrestrial planets.  Orbital migration has been proposed as a potential avenue for isolating Mercury's orbit \citep{broz21,clement21_merc_outwardmig}, however this idea has not been robustly tested in conjunction with the impact scenario of \citet{asphaug14}.  It is also possible that Mercury's mantle was incrementally tidally stripped as the result of repeated close encounters with Venus \citep{deng20,fang20}, however this scenario carries with it many of the same drawbacks as models requiring an actual collision.

Given the amount of mantle material that must be redistributed to replicate Mercury's modern CMF, the collisional velocities involved in either scenario are necessarily high (approaching six times the mutual escape velocity for oblique impacts).  Such extreme events are outside of the spectrum of collisional outcomes that occur in transitional N-body studies of terrestrial planet formation \citep{jackson18,clement19_merc}.  However, the region around Mercury's modern orbit is also home to a number of strong orbital resonances \citep{michelandfroeschle97,batygin15} that can drive chaotic evolution of proto-planets in the region and even trigger extremely high-velocity impacts \citep{clement21_merc_lone_survivor}.  Thus, if a chain of embryos or planets formed inside the modern orbit of Mercury \citep[in contrast to how terrestrial planet formation models typically treat the region,][]{volk15}, energetic, mantle-stripping events might be quite common.  However, studies of this scenario tend to form Mercury at extremely low semi-major axes \citep{clement21_merc_lone_survivor}.  Excessive velocities can also be avoided if Mercury's mantle is stripped incrementally through a series of multiple hit-and-run events \citep{chau18,franco22}.  Indeed, \hbindex{N-body simulation}s incorporating collisional fragmentation do find that high-CMF planets can form with somewhat regularity as a result of fortuitous chains of imperfect accretion events that happen to transpire during the highly stochastic giant impact phase \citep{chambers13,clement19_frag,clement23}.  This possibility disentangles Mercury's mass and CMF from its specific location in the Solar System by arguing that Mercury-like planets should be rather common in the cosmos.  Indeed, exoplanet surveys have recently begun to reveal a number of Super-Earths with densities that might be indicative of a Mercury-like internal structure \citep{rodriguez23,mah23}.

Aside from the aforementioned challenges related to replicating Mercury's dynamical offset from Venus, perhaps the largest problem for chaotic hypotheses to overcome is the high likelihood of debris and disrupted volatile re-accretion \citep{benz07,gladmanandcoffey09,crespi21}.  As hydrodynamical simulations typically find that a large quantity of the material ejected from giant impacts is in the form of dust, it is possible that the majority of the ground-up mantle material that was removed via \hbindex{Poynting–Robertson drag} \citep{melis12} or interactions with the young Sun's strong wind \citep{spaldingandadams20}.

\subsubsection{Orderly Models}

Early models attempting to explain Mercury's curious composition and internal structure argued that silicates in the region around proto-Mercury were evaporated as a result of heightened stellar activity during the pre-Main Sequence phase \citep{morgan80}.  However, such a scenario is inconsistent with \hbindex{MESSENGER}'s findings that less volatile, lithophile elements in Mercury's crust have roughly chondritic abundances.  While a number of models for preferential iron-enrichment of Mercury's planetesimal precursors have been proposed since \hbindex{MESSENGER}, it is important to note that these scenarios axiomatically require Mercury grow within a separate reservoir of material than Earth and Venus.  In order to match constraints for Mercury's mass and orbit, this innermost component of the inner disk must be significantly depleted in mass relative to the outer disk component, and also posses a distinctive surface density profile \citep[equation 11:][]{lykawakaito17,clement21_merc_insitu,lykawka23}.  While such a structure is necessarily contrived and inconsistent with planetesimal formation and accretion models \citep{boley14}, convergent \citep{broz21} or outward \citep{clement21_merc_outwardmig} migration of terrestrial embryos have been proposed as dynamical mechanisms for re-sculpting the inner component of the terrestrial disk.

One process potentially responsible for iron enrichment of Mercury's planetesimal precursors is the \hbindex{photophoretic effect} \citep{wurm2013}.  When $\sim$mm-scale particles in a gaseous disk are exposed to non-isotropic solar radiation, they can migrate substantially.  The extent of this migration varies based off the local thermal gradient and the size of the particles themselves \citep{krauss05}.  Through this process particles can be size sorted, although it is still unclear how efficient this mechanism is in the opaque mid-plane of \hbindex{protoplanetary disk}s \citep{cuzzi08}.  Nevertheless, advanced disk chemistry models incorporating effects such as this have demonstrated the formation of a compositionally distinct inner region of the terrestrial disk \citep[e.g.][]{moritary14,pignatale16}.

Magnetic aggregation has also been proposed as a mechanism for iron-enrichment of planetesimals in the inner terrestrial disk \citep{kruss20}.  Laboratory experiments have shown that suspended magnetized particles within ferromagnetic fluids generate chain-like structures as a result of their behaving like dipoles \citep{kruss18,jungmann22}.  If the magnetic field in the region of the disk around proto-Mercury is sufficiently strong, it is possible that these long iron-rich chains can be preferentially incorporated into planetesimals formed via the \hbindex{streaming instability}. 

Another idea for altering the chemistry of planetesimals in the neighborhood of the forming Mercury comes from high temperature experiments with pre-solar interplanatary dust particles in carbon-rich, oxygen-depleted environments.  \citet{ebelandalexander11} showed that these conditions favor the formation of condensates with Fe/Si ratios half that of the bulk Mercury.  Indeed, anhydrous chondritic interplanatary dust particles have Carbon abundances that are an order of magnitude more than those of CI chondrites \citep{ebelandalexander02}.  However, the efficiency of this mechanism is highly dependent on the unconstrained chemical structure of the disk mid-plane in the vicinity of proto-Mercury.

Recently, \citet{johansen22} developed a model that relies on the high surface tension of iron \citep{ozawa11} relative to silicates \citep{lummenandkraska05}.  Iron particles nucleate homogeneously only under very supersaturated conditions, thus promoting the depositional growth of a small population of nucleated iron particles embedded in a large distribution of iron pebbles.  Contrarily, while silicates nucleate under similar conditions they also obtain smaller sizes than iron particles \citep{kashchiev06}.  This dichotomy in turn promotes the growth iron-rich, silicate-poor planetesimals via the \hbindex{streaming instability} \citep{youdingoodman05}.

Substantial investigation is still required to better understand whether any of these ``orderly'' enrichment processes played a role in Mercury's growth, and if so to what degree.  While more sophisticated planetesimal formation, disk chemistry and dynamical formation models will surely shed additional light on the potential source of Mercury's high CMF, the subsequent section details ways in which exoplanet demographics might help break degeneracies between the ``orderly'' and ``chaotic'' hypotheses in the future.

\subsection{Constraining and distinguishing formation scenarios}

A number of models for the late stage accretion of terrestrial planets in the Solar System have been proposed over the past decade.  Among others, recently proposed models that still still require additional study include the convergent migration scenario of \citet{broz21}, the \hbindex{pebble accretion} models of \citet{levisonetal15pnas} and \citet{johansen21}, and the chaotic excitation scheme proposed by \cite{lykawka23}. This review focused on three specific models that are not only consistent with the masses and orbits of the terrestrial planets (\hbindex{RMC} and \hbindex{AMD}), the structure of the asteroid belt and the origins of water in the inner Solar System \citep{walshetal11,clement18,izidoro22_natast}; but perhaps more importantly are self-consistent with envisioned dynamical \citep{tsiganisetal05,morbidellicrida07,nesvornymorbidelli12} and cosmochemical \citep{kleineetal09,dauphaspourmand11,kruijeretal17,dauphas17} models of global Solar System evolution.  While numerical simulations of each of these models seems to produce ``good'' terrestrial analogs systems with similar regularity, only one of them is potentially correct. So how can we hope to distinguish between them? 

Empirical tests to discriminate between these models in the future may be based on space observations of Solar System minor bodies \citep{morbyraymond16}, or high precision isotopic measurements of different planetary objects \citep[e.g.][]{tang14,dauphas17,dauphas24}. On the theoretical side, models may be ruled out or bolstered by more sophisticated numerical simulations with superior treatments of various physical mechanisms like \hbindex{pebble accretion} and planetary migration \cite{izidoroetal16}, or through approaches that blend dynamical and disk chemistry models \citep{lichtenberg21,morby21,izidoro22_natast}. 

The \hbindex{Grand Tack} model requires a very specific large-scale giant planet migration  sequence. One of the main loose ends of the \hbindex{Grand Tack} is that it is not clear if the required inward-then-outward large scale migration is possible when gas accretion onto Jupiter and Saturn is self-consistently computed \citep{raymondmorbidelli14}. Unfortunately, our understanding of gas accretion onto cores is still incomplete. Indeed, hydrodynamical simulations of the growth and migration of Jupiter and Saturn typically invoke a series of simplifications considering the challenge in performing high-resolution self-consistent simulations of this process \citep[e.g.][]{zhangzhou10,pierensetal14}.

The major drawback of the \hbindex{Early Instability} model is the very specific timing \citep[$<$10 Myr, and possibly within the limited window of 1-5 Myr as found in][]{clement19_frag,clement21_earlyinst}.  Indeed, while many Solar System constraints pin the instability's occurrence down to the first 100 Myr after disk dispersal, a number of these \citep[e.g.][]{nesvorny15a,morby18,nesvorny23_craters} fit in best with an instability that occurs in the latter half of this window.  Moreover, the Earth's atmospheric inventory of Xenon appears to have a cometary signature that is missing in the mantle \citep{marty17}; suggesting a substantial delivery of volatiles from comets (presumably dispersed via the instability) \textit{after} core closure.  However, recent work by \citet{joiret23} found that cometary bombardment can occur over a more prolonged window of tens of Myr after the onset of the instability.  Therefore, it is at least plausible that the instability could have occurred quite early and still delivered cometary Xenon to Earth well after its formation \citep[note, there are dynamical reasons to favor an extremly \hbindex{Early Instability} as well][]{quarles19,liu22}.

The Achilles' heel of the rings model lies in the assumed initial condition of primordial pressure bumps.  If no such structure existed in the Solar System, or planetesimal formation was in fact efficient beyond 1-1.5 au, another mechanism must have been responsible for limiting Mars' mass \citep{morbyraymond16}.  Moreover, while the proposition of highly-localized planetesimal formation \citep{drazkowska16,raymondizidoro17b,lichtenberg21} is indeed a powerful explanation for the \hbindex{NC/CC dichotomy} \citep{kruijeretal17}, it is worth considering the diversity of ages and compositions \citep[see, for example:][]{alexander22} of the various members of the highly diverse set of sub-classes within each major chondrite groups.  Moreover, certain ungrouped \hbindex{meteorites} have been interpreted to evidence very early mixing of NC and CC-derived materials \citep{spitzer22}.  Thus, from a meteroites perspective, the full story of planetesimal formation is likely to be far more complicated than assumed in any terrestrial planet formation model.

%10/23 MATT: I think that Andre's new paper resolves this somewhat.  But we could still use this text in the rings section
%Additionally, it has been  suggested that at least two generations of planetesimal were born in the inner Solar System. The oldest population is associated with the parent bodies of iron \hbindex{meteorites} which formed around half-million years after \hbindex{CAIs} \citep{kruijeretal12}. The youngest one is associated with chondritic \hbindex{meteorites}, and formed after $\sim$3~Myr \cite{villeneuveetal09}. The late formation of asteroids (potentially after the short-lived radionuclides such as $^{26}$Al became inefficient as a heat source) is also supported by the fact that \hbindex{S-type} asteroids are dominated by thermally undifferentiated bodies \citep{weisselkinstanto13,scheinbergetal15}. If the terrestrial planets formed from an earlier generation of planetesimals than the \hbindex{S-type}s, this would conflict with the \hbindex{Planet Formation from Rings} model, in which the two populations are drawn from the same source.

Strong tests aiming to disentangle these models may also emerge from more complex multidisciplinary approaches combining the accretion history of Earth produced in \hbindex{N-body simulation}s with models of core-mantle differentiation and geochemical models \citep[see, for example:][]{rubieetal15,rubie16,fischer17,zube19,gu23}. Analyses such as these may eventually be the key to distinguishing between the various terrestrial planet formation scenarios.

Exoplanet demographics can also provide valuable insights into the spectrum of processes that potentially played roles in the formation of our own Solar System.  For instance, additional ``\hbindex{super-Mercuries}'' will undoubtedly be discovered in the near future through more precise density measurements facilitated by next generation telescopes.  If such planets occur with similar regularity at different positions in multi-planet systems, it would support the idea that Mercury formed as the result of a series of mantle-stripping impacts \citep{mah23}.  Contrarily, if \hbindex{super-Mercuries} are over-represented in the subset of planets that are the closest to their system's central star, it would appear more likely that Mercury formed from a population of iron-rich planetesimals.

\section{Terrestrial planet formation in the context of exoplanets}

If we understand terrestrial planet formation in the Solar System (at least to some degree), then we can hopefully extrapolate this knowledge to terrestrial planet formation in a more general setting.  The thousands of known exoplanets -- many of which are close to Earth-sized -- offer an excellent testbed for our models. The difficulty is in knowing which exoplanets are truly analogous to our own terrestrial planets and which are entirely different beasts.

As of 2023 there are more than 5,000 confirmed exoplanets~\citep{christiansen22}. The bulk were discovered either by radial velocity surveys using Doppler spectroscopy~\citep{fischer14} or by transit surveys; notably NASA's {\it \hbindex{Kepler}} mission~\citep{borucki10}. It is now know that at least a few percent (and perhaps even $\gtrsim$10$\%$) of Sun-like stars host gas giant planets~\citep{mayor11,howard12,petigura18,gan22,beleznay22,bryant23} but that hot Jupiters exist around only $\sim$1\%~\citep{wright12,howard10}. While most gas giants are found on orbits beyond 0.5-1 AU~\citep{butler06,udry07b}, the population is dominated by planets on eccentric orbits.  True Jupiter `analogs' -- with orbital radii larger 2 AU and eccentricities smaller than 0.1 -- exist around only $\sim$1\% of stars like the Sun~\citep{martin15,morbyraymond16}. This is thus an upper limit on the occurrence rate of potential true Solar System analogs.  While analogs of individual planets like Venus or Uranus may be more common, bulk system architectures like ours cannot be.

At first glance, "\hbindex{hot super-Earth}s'' and ``\hbindex{sub-Neptune}s'' -- often defined as being smaller than $4 \rearth$ or $20 \mearth$ with orbits shorter than $\sim$100 days -- seem tantilizing similar to the solar system's terrestrial planets.  Super-Earths have been shown to orbit at least half of all main sequence stars, including both Sun-like~\citep{mayor11,howard12,fressin13,petigura13,zhu18} and low-mass stars~\citep{bonfils13,mulders15b}. Many systems have been found to host multiple super-Earths.  These so-called ``multiples'' tend to have compact orbital configurations and similar-sized planets~\citep{lissauer11,lissauer11b,millholland17,weiss18}. 
 Extensive radial velocity monitoring of {\it \hbindex{Kepler}} super-Earths has also revealed a noteworthy dichotomy: smaller planets (radii $\lesssim$1.5 $\rearth$) have high densities and are indeed rocky (super-Earths) whereas larger planets (\hbindex{sub-Neptune}s with 1.5 $\lesssim R \lesssim$ 4 $\rearth$) tend to have lower densities and likely more like the solar system's ice giants than its terrestrial planets \citep{weiss13,marcy14,weiss14}.  The division between super-Earths and mini-Neptunes appears to lie close to $1.5 \rearth$~\citep{weiss14,lopez14,rogers15,wolfgang16,chen17}.  

This section begins with a discussion of various models for the origin of super-Earths (broadly-defined to include all planets smaller than $4 \rearth$), and then concludes with an overview of how terrestrial planet formation may proceed in systems with gas giants on orbits very different from Jupiter's.

\subsection{Origin of super-Earth systems}

The observed population of super-Earths is rich enough to provide quantitative constraints on formation models:
\begin{enumerate}
\item Their occurrence rate~\citep[$\sim 50\%$ around main sequence stars;][]{mayor11,fressin13,petigura13,dong13}. 
\item Their multiplicity distribution. Indeed, it is easier to precisely determine the masses and orbits of planets in systems with multiple planets, for instance through transit-timing variations offer \citep{lissauer11b}. However, the observed population of exoplanets has more single super-Earth systems than multiple systems, which is sometimes referred to as the ``\hbindex{Kepler dichotomy}''~\citep{fang12,tremaine12,johansen12}.
\item The distribution of of orbital period ratios of adjacent planets in multiple planet systems~\citep{lissauer11b,fabrycky14}. Of particular curiosity is the fact that the distribution is not preferentially peaked at first order MMRs (as Type-1 \hbindex{migration model}s would predict) \textit{and} is also not particularly uniform \citep{fabrycky14}.
\item The division between rocky super-Earths and gas-rich mini-Neptunes at $\sim 1.5 \rearth$ ~\citep{weiss14,lopez14,rogers15,wolfgang16,chen17}.
\end{enumerate}

Many models have been proposed to explain the origin of super-Earths.  Before almost any super-Earths were known, \cite{raymond08} proposed six potential formation pathways for their formation and laid out a simple framework to use observations of system architecture and planet bulk density to differentiate between them.  Several of those pathways were quickly disproven because they did not match observations; for instance, one mechanism proposed that super-Earths form from material shepherded inward by a migrating giant planet~\citep{fogg05,fogg07,raymond06b,mandell07}.  It was quickly shown that there is no correlation between close-in gas giants and super-Earths -- to the contrary, there is generally an anti-correlation between hot Jupiters and other close-in planets~\citep{latham11,steffen12}.

At the time of the writing of this chapter, two models remain viable: the {\it migration} and {\it drift} models.  Nevertheless, it is still worth explaining why simple, in-situ growth of super-Earths is not a viable formation mechanism.  In-situ growth of super-Earths was first proposed by \cite{raymond08}, and immediately discarded because of the prohibitively large disk masses required. It was re-proposed by \cite{hansen12,hansen13} and \cite{chiang13}, and was again refuted for both dynamical and disk-related reasons~\citep{raymond14,schlichting14,schlaufman14,inamdar15,ogihara15}. The simplest argument against in-situ growth is as follows. If super-Earths form in-situ then they must grow extremely quickly because in extremely dense disks possessing many Earth-masses worth of material extremely close to the central star.  However, if planets form that quickly in massive gas disks, they must migrate.  In fact, the disks required to build super-Earths close-in are so dense that aerodynamic drag acts on full-grown planets on a shorter timescale than the disk dissipation timescale~\citep{inamdar15}. Thus, in-situ growth implies that planets must migrate.  If they migrate then their orbits change and thus they cannot truly form ``in-situ''.

The \hbindex{drift model} proposes that planets form sequentially in concentric, gravitationally unstable rings that form as dust drifts inward.  Dust is indeed expected to coagulate and drift inward~\citep[e.g.][]{birnstiel12} and if there exists a trap in its route towards the central star, a fraction of the mass in drifting pebbles can be captured.  \cite{chatterjee14} proposed that, once a pebble ring attains a high enough density, it can directly collapse into a full-sized planet.  The inner edge of the dead-zone (the region where collapse can occur) subsequently retreats, thus shifting the formation location of the next super-Earth.  This model is promising and the subject of a series of papers~\citep{chatterjee14,chatterjee15,boley14,hu16,hu17}.  However, additional work is still required to comprehensively address each of the constraints listed above. 

\begin{figure}
% For example, with the graphicx style use
\centering
\includegraphics[scale=0.5]{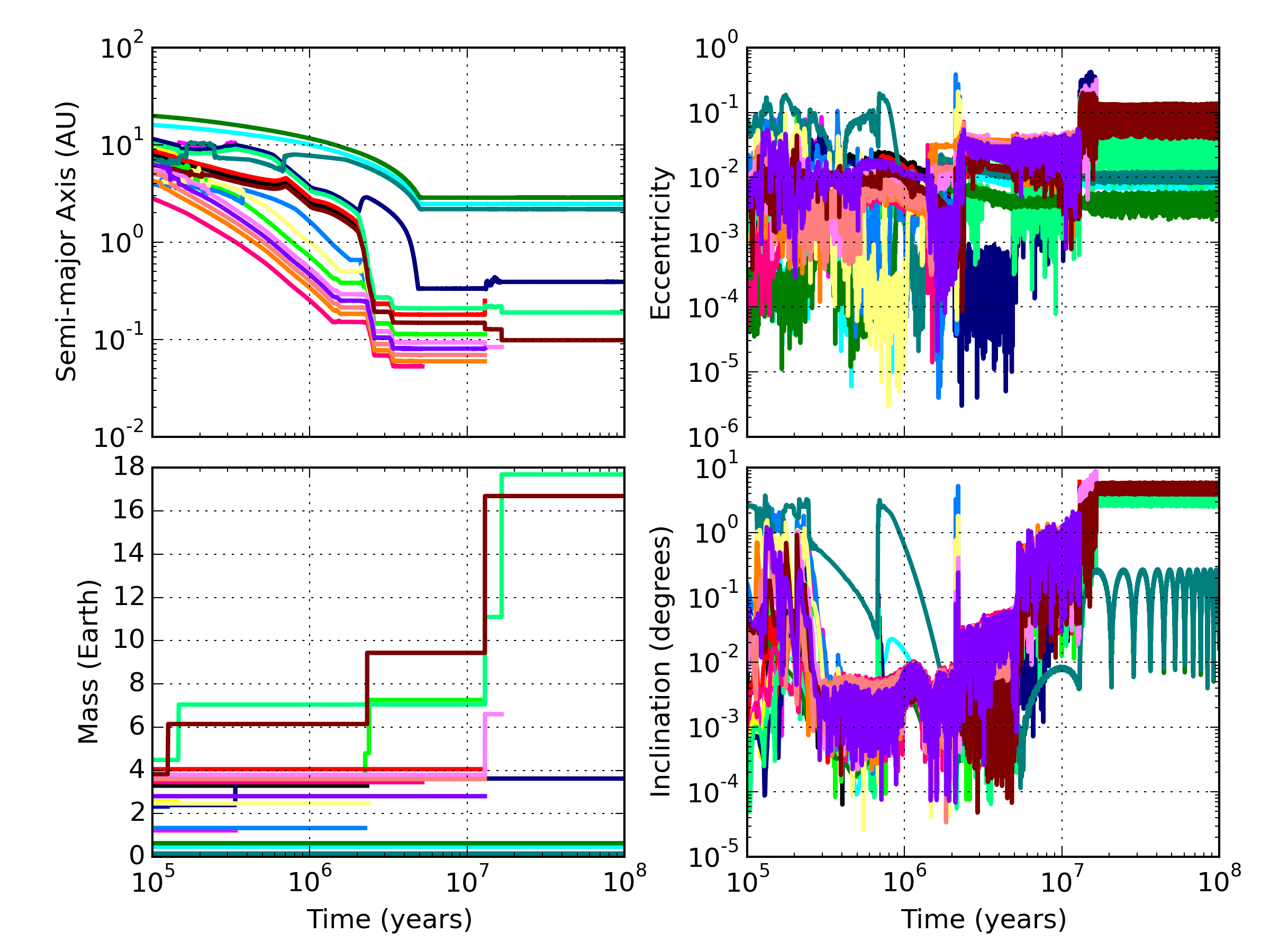}
\caption{Mass and orbital evolution of a system forming close-in super-Earths, reproduced from \cite{izidoro17}.  A set of $\sim$Earth-mass planetary embryos are initialized beyond the \hbindex{snowline} and subsequently migrate inward (undergoing occasional collisions as they do) to create a long resonant chain.  In the plotted case, 10 super-Earths initially reside in a chain of resonances interior to 0.5 AU. When the gas disk dissipates at 5 Myr, the system remained quasi-stable for a few Myr before eventually undergoing a large scale instability that leads to a phase of late collisions. The final system consists of just three (relatively massive) super-Earths with modest eccentricities and a large enough mutual inclinations to preclude the transit detection of all three planets.}
\label{fig:migration} 
\end{figure}

The \hbindex{migration model} proposes that planetary embryos grow far from the central star, and subsequently move inward via \hbindex{Type-I migration} ~\citep[see example in Fig.~\ref{fig:migration};][]{goldreich80,ward86,tanaka02}. Given that disks have magnetically-truncated inner edges~\citep[e.g.][]{romanova03,romanova04}, inward migrating embryos may be caught at so-called planet traps \citep{lyraetal10,hasegawapudritz11,hasegawapudritz12,hornetal12,bitschetal14,alessietal17} before eventually reaching the inner edge where a strong torque prevents them from falling onto the star~\citep{masset06}.  Systems of migrating embryos thus pile up into chains of \hbindex{mean motion resonance}s anchored at the inner edge of the disk~\citep{cresswell07,terquem07,ogihara09,mcneil10,cossou14,izidoro14}.  While collisions are common during this phase, they only temporarily destabilize the resonant chain before it is quickly reconstituted. Short lived gaseous disks \citep[e.g.][]{hasegawapudritz11,bitsch15,alessietal17} or reduced gas accretion rates \citep{lambrechtslega17} have been proposed as potential mechanisms for preventing these embryos from growing further and becoming gas giant planets before they have a chance to migrate inward. When the disk dissipates, eccentricities and inclinations of the resonant planets are no longer damped~\citep{tanaka04,cresswell07,bitsch10}.  This causes most resonant chains to become unstable, and triggers a late phase of giant collisions in a gas-free (or at least, very low gas density) environment~\citep{terquem07,ogihara09,cossou14,izidoro17,izidoro21_super_earth}. Assuming that 5-10\% of systems remain stable after the disk dissipates, the surviving systems in simulations of this process provide a quantitative match to both the observed super-Earth period ratio and multiplicity distributions~\citep{izidoro17,izidoro21_super_earth}. In this manner, the \hbindex{Kepler dichotomy} is an observational artifact generated by the bimodal inclination distribution of super-Earths, a few of which have very low mutual inclinations (and thus a high probability of being discovered as multiple systems), however the majority have significant mutual inclinations that are the consequence of these late instabilities~\citep{izidoro17}.  The model also neccessarily explains the existence of super-Earths in resonant chains like \hbindex{Kepler}-223 \citep{mills16}.

How can we hope to use observations to differentiate between the drift and \hbindex{migration model}s?  In its current form the \hbindex{migration model} is built on the assumption that embryos large enough to migrate should preferentially form far from their stars, past the \hbindex{snowline}. In the Solar System it is indeed thought that large embryos formed in the outer Solar System and became the cores of the giant planets, whereas small embryos formed in the inner Solar System and became the building blocks of the terrestrial planets~\citep{morby15}.  Given their distant formation zones, the \hbindex{migration model} thus predicts that super-Earths should be predominantly water-rich with correspondingly low densities \citep{raymond08}.  In contrast, inward migrating pebbles in the \hbindex{drift model} should have time to devolatilize before they collapse into proto-planets, and the resulting accretion super-Earths should be predominantly rocky.  However, this difference depends strongly on where the first planetesimals form, as these serve as the seeds for embryo growth .  This question is unresolved: some studies find that planetesimals first form at $\sim$1 AU~\citep{drazkowska16} whereas others find that planetesimals first form past the \hbindex{snowline}~\citep{armitage16,drazkowskaalibert17,carreraetal17}.

The transition between super-Earths and \hbindex{sub-Neptune}s is thought to be a result of a competition between accretion and erosion~\citep[][see also chapter by Schlichting]{ginzburg16,lee16}. Growing planetary embryos accrete primitive atmospheres from the disk~\citep[e.g.][]{lee14,inamdar15}, but accretion is slowed by heating associated with small impacts~\citep{hubickyj05}, and also eroded by large impacts~\citep{inamdar16} during the disk's dissipation~\citep{ikoma12,ginzburg16}.  Photo-evaporation of close-in planets may also play an important role by preferentially stripping the atmospheres of low-mass, highly-irradiated planets~\citep{owen13,owen17,lopez16}. Planets located in the ``\hbindex{photo-evaporation valley}'' -- the region of very close-in orbits where any atmospheres should have been stripped from low-mass planets -- appear to be mostly rocky~\citep{lopez17,jin18}.  Of course, this only applies for the closest-in planets, which can be plausibly built from rocky material shepherded inward by migrating, volatile-rich planets in the \hbindex{migration model} \citep{izidoro14}. However, it is extremely difficult to accurately determine the compositions of more distant planets because the can be derived from at least three different types of building blocks: rock, water and Hydrogen~\citep{selsis07,adams08}.  Thus, only the most extreme densities can provide useful insights (e.g., very high density planets are likely to have little water or Hydrogen).

%The role of the central star remains to be fully incorporated into models of super-Earth formation.  Compared with FGK stars, {\it \hbindex{Kepler}} found that M stars have more super-Earths and fewer mini-Neptunes, and for a higher total planet mass~\citep{mulders15a,mulders15b}.  This remains to be clearly understood, and may be linked with the low abundance of gas giants around M stars~\citep{lovis07,johnson07}.

\subsection{Terrestrial planet forming in systems with giant exoplanets}

The dynamical evolution of such systems is thought to be quite different than that of Jupiter and Saturn. Indeed, the median eccentricity of giant exoplanets is 0.25~\citep{butler06,udry07b}, five times larger than that of Jupiter and Saturn.  Although observational biases preclude a clear determination, most giant exoplanets are also located somewhat closer to their stars than the solar system's gas giants \citep[typically at 1-2 AU:][]{cumming08,mayor11,rowan16,wittenmyer16}.

Two key processes are thought to be responsible for shaping the orbital distribution of giant exoplanets: type II inward migration and planet-planet scattering. While Jupiter and Saturn certainly migrated, the extent of migration remains unclear.  Similarly, a wide range of migration schemes are possible for exoplanets.  Indeed, some may have migrated all the way to the inner edge of the disk to become hot Jupiters~\citep{lin86,lin96}.  The high eccentricities of giant exoplanets are easily explained if the observed planets are the survivors of system-wide instabilities during which giant planets scattered repeatedly off of each other during close passages inside each others' Hill spheres~\citep{rasio96,weidenschilling96,lin97,adams03,moorhead05,ford03,chatterjee08,ford08,raymond08,raymond10}.  This phase of planet-planet scattering typically concludes with the ejection of one of more planets. In some cases scattering can push planets to sufficiently high eccentricities that they pass very close to their stars at pericenter; thus allowing tidal dissipation to circularize and shrink their orbits.  This has been proposed as an alternate channel for the formation of hot Jupiters~\citep{nagasawa08,beauge12}.

It is natural to question how giant planet migration and scattering affect the growth and evolution of potential terrestrial planets.  Giant planet migration has been shown to be much less destructive to would-be terrestrial planets than was generally assumed in the late 1990s and early 2000s~\citep{gonzalez01,lineweaver04}.  An inward-migrating gas giant does not simply collide with the material in its path~\citep[except in rare circumstances;][]{tanaka99}.  Rather, strong inner \hbindex{mean motion resonance}s acting in concert with \hbindex{gas drag} shepherd material inward, often helping to catalyze the formation of planets interior to the giant planets' final orbits~\citep{fogg05,fogg07,raymond06b,mandell07}.  Moreover, a significant amount of the material undergoing close encounters with the migrating giant planet is scattered outward and stranded on eccentric and inclined orbits.  This material can subsequently re-accumulate into a generation of terrestrial planets that tend to have extremely wide feeding zones and are thus very volatile-rich~\citep{raymond06b,mandell07}. 

The eccentricity distribution of eccentric giant planets can be matched by planet-planet scattering models \citep[e.g.][]{chatterjeeetal08,jurictremaine08,fordrasio08,raymondetal08}. 
In contrast to giant planet migration, giant planet scattering is typically very harmful to the process of terrestrial planet formation.  When gas giants destabilize they scatter each other onto eccentric orbits, and any small bodies (planetesimals, planetary embryos or planets) in their path are typically destroyed~\citep{veras05,veras06,raymond11,raymond12,matsumura13,marzari14,carrera16}.  Objects that are closer-in than the gas giants are preferentially driven onto such eccentric orbits that they collide with the host star, whereas more distant objects are typically ejected~\citep{raymond11,raymond18,marzari14}.

\section{Putting our Solar System in context}

How can we understand our Solar System in a larger context?  What are the key processes that make our system different than most? 

Jupiter is likely the Solar System's primary architect. Let us consider its potential effects on the greater system at different phases of its growth. Jupiter's core was perhaps seeded by an early generation of planetesimals that then grew by \hbindex{pebble accretion}~\citep[][]{ormeletal10,lambrechts12,lambrechts14,chambers21}.  It is unclear \textit{where} this took place. Studies have covered the full spectrum of possibilities, from distant formation followed by inward migration~\citep{bitsch15,deienno22} to in-situ growth~\citep{levisonetal15nat} to close-in formation followed by outward migration~\citep{raymond16}.  Nonetheless, once its core reached $\sim 20 \mearth$ it likely created a pressure bump exterior to its orbit that blocked the inward pebble flux~\citep{lambrechtsetal14,kruijeretal17}. This acted to starve the inner Solar System of the raw materials needed to form super-Earths or giant planets \citep{morby15,lambrechts19,izidoro22_natast,chambers23}. 

Although the direction, duration and speed are uncertain, Jupiter's core subsequently migrated; thereby shepherding any nearby cores and planetesimals~\citep{izidoro14}. When Jupiter underwent rapid gas accretion it strongly perturbed the orbits nearby small bodies, scattering them across the Solar System~\citep[and implanting some in the inner Solar System][]{raymondizidoro17a}.  It carved a gap in the disk and transitioned to slower, \hbindex{Type-II migration}~\citep{lin86,ward97,crida06}.  Jupiter now served as a strong barrier for more distant planetary embryos that would otherwise migrate inward to become close-in super-Earths~\citep{terquem07,ogihara09,izidoro15}.  Blocked by Jupiter and Saturn, these embryos instead continued to accrete embryos in-place and became the ice giants~\citep{jakubik12,izidoro15b,chau21}.  Once the disk dissipated, Jupiter's dynamical influence played a key role in the late-stage accretion of the terrestrial planets and the dynamical sculpting of the asteroid belt~\citep[particularly during the epoch of giant planet instability,][]{raymondetal14,deienno18,clement18}.

There are thus two potential ways that Jupiter may explain why the Solar System is different, specifically our lack of super-Earths.  The first is by blocking the pebble flux and starving the growing terrestrial planetary embryos. The second is by blocking the inward migration of large cores~\citep{izidoro15b}.  

%If the Solar System's presumed primordial super-Earths formed by migration, they must have migrated inward {\it through} the building blocks of the terrestrial planets. \hbindex{Type-I migration} may be directed inwards or outwards (see chapter by Nelson) or even be halted depending on the disk local properties \citep{ward86,ward97a,paardekoopermellema06,paardekoopermellema08,baruteaumasset08,paardekooperpapaloizou08,kleyetal09,kleycrida08,paardekooperetal10,paardekooperetal11}. However, as the disk evolves and cools down any type-I migrating planets are eventually released to migrate inwards \citep{lyraetal10,hornetal12,bitschetal14}. If their migration was slow, the super-Earths would have swept the region around 1 AU clean of rocky material such that any planets that formed there would be decidedly un-Earth-like~\citep[see right-panel of Figure \ref{fig:migratingSE}][]{izidoro14}.  However, if their migration was fast, super-Earths would simply migrate past rocky planetary embryos without completely disrupting their distribution (see left-panel of Figure \ref{fig:migratingSE}). Alternately, if super-Earths formed by the \hbindex{drift model}, it is plausible that they could have accumulated material close-in without perturbing the terrestrial planets' growth.  Thus, the growth of a population of close-in super-Earths in the Solar System seems plausible.

\begin{figure}
\hspace{-1.3cm}
\includegraphics[scale=.5]{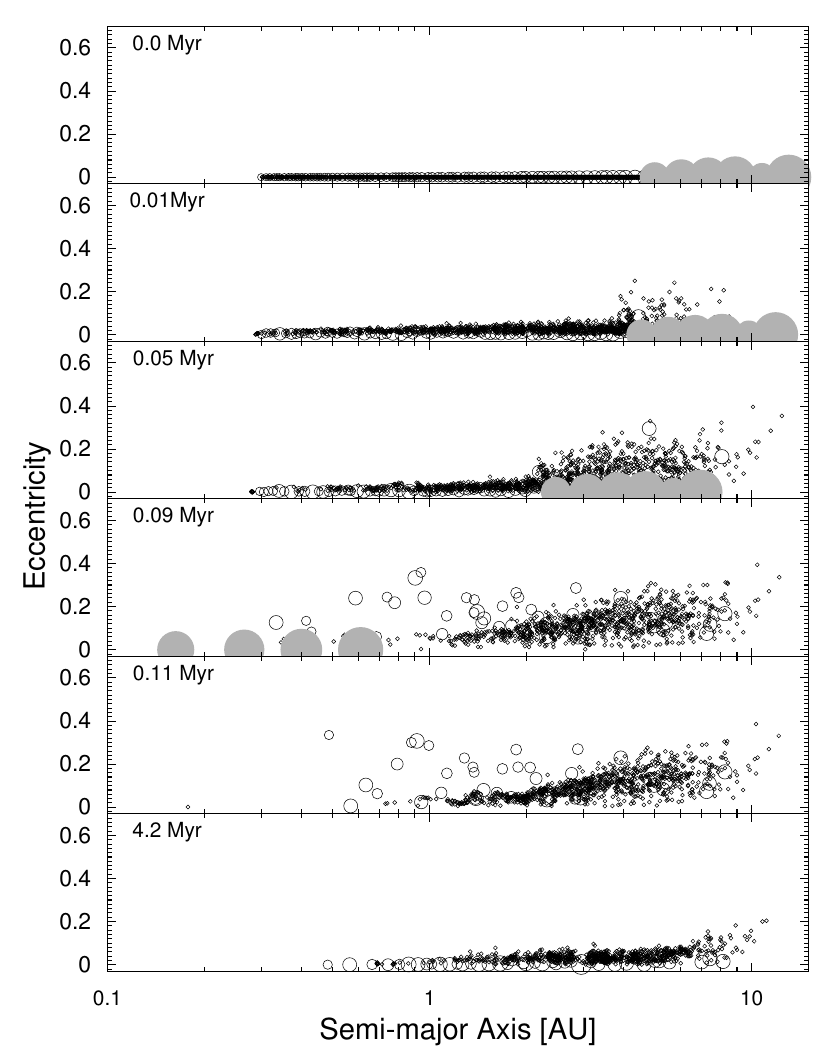}
\includegraphics[scale=.5]{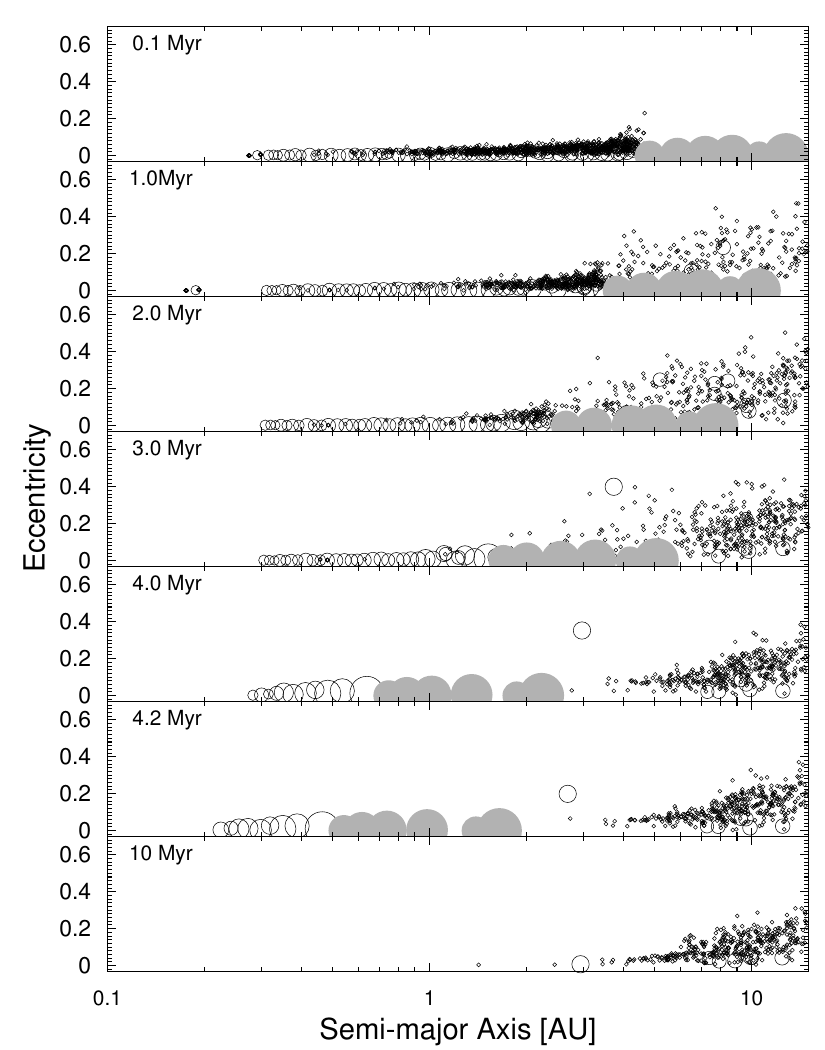}
\caption{Snapshots of the dynamical evolution of a population of planetesimal and planetary embryos in the presence of migrating super-Earths. The gray filled circles represent the super-Earths. Planetary embryos and planetesimals are shown by open circles and small dots, respectively. The super-Earth system is composed of six super-Earths  with masses roughly similar to those of the \hbindex{Kepler} 11 system \citep[e.g.][]{lissaueretal13}. The left-panel shows a simulation where the system of super-Earths migrate fast, in a short timescale of about 100~kyr. The right-panel represents a simulation where the system of super-Earths migrate slowly, in a timescale comparable to the disk lifetime. Figure adapted from \citep{izidoro14}}
\label{fig:migratingSE}
\end{figure}

%The next question is: what could have happened to such a population of close-in super-Earths?  \cite{volk15} proposed that they were cataclysmically ground to dust by a series of giant erosive collisions.  \cite{batygin15} proposed instead that Jupiter's migration led to a spike of collisional grinding at $\sim$1AU that produced a population of small ($\sim$100-m) planetesimals that drifted inward and became trapped in exterior resonances with the super-Earths.  Strong aerodynamic dissipation in the planetesimals' orbits subsequently pushed the super-Earths onto the young Sun.

%While provocative, each of these studies neglects the fact that planet-forming disks have inner edges~\citep[e.g., see magneto-hydrodynamic simulations of][]{romanova03,romanova08}. These edges prevent planets or debris from simply falling onto the Sun~\citep[see discussion in][]{raymond16}. Rather, all studies to date suggest that the processes that generate dust or debris should rather catalyze the further growth of close-in planets~\citep{leinhardt09,kenyon09,chatterjee14}.  If super-Earths indeed formed in the Solar System, they should still be here. Thus, there is currently no compelling evidence that super-Earths ever formed in the Solar System.

Considering only the Sun and Jupiter, exoplanet statistics tell us that the Solar System is already at best a 1\% outlier~\citep[and more like 0.1\% when considering all stellar types:][]{petigura18,gan22,beleznay22,bryant23}.  Yet it is likely that Earth-sized planets on Earth-like orbits may be far more common.  The Drake equation parameter eta-Earth -- the fraction of stars that host a roughly Earth-mass or Earth-sized planet in the habitable zone -- has been directly measured for low-mass stars to be tens of percent~\citep{bonfils13,kopparapu13,dressing15,bergsten23}.  Yet how `Earth-like' are such planets?  Without Jupiter, would a planet at Earth's distance still look like our own Earth? 

When viewed through the lens of planet formation, two of Earth's characteristics are unusual: its water content and formation timescale.  The building blocks of planets tend to either be very dry (possessing only a fraction of a percent of water like non-carbonaceous chondrites) or very wet ($\sim$10\% water like carbonaceous chondrites, or $\sim$50\% water like comets).  In contrast, Earth's composition can be explained by having grown mostly from dry material with only a sprinkling of wet material \citep{marty12,dauphas17,dauphas24}.  A simple explanation for this mixture is that, even though Jupiter's formation provided a sprinkling of water-rich material~\citep{raymondizidoro17a}, once the giant planets grew large enough the blocked all subsequent water delivery \cite[e.g.][]{morbidellietal16,satoetal16}.  Without Jupiter it stands to reason that Earth should either be completely dry or, more likely, much wetter. 

Earth's last giant impact is constrained not to have happened earlier than $\sim$40 Myr after \hbindex{CAIs}~\citep{toubouletal07,kleineetal09,aviceetal17}. However, most `Earth-like' planets probably form much faster.  Super-Earths typically complete their formation shortly after dispersal of the gaseous disk~\citep{izidoro17,alessietal17}. Accretion in the terrestrial planet zone of low-mass stars is similarly fast whether or not migration is accounted for~\citep{raymond07,lissauer07,ogihara09,clement22}. The geophysical consequences of fast accretion remain to be further explored, but it stands to reason that fast-growing planets are likely to be hotter and may thus lose more of their water compared with slower-growing planets like Earth.  This could in principle counteract our previous assertion that most terrestrial planets should be wetter than Earth.

While other Earths remain a glamorous target for exoplanet searches, developing a more comprehensive understanding of the ways in which exoplanets are similar to, and different than the worlds of our own Solar System is also an extremely compelling line of inquiry.  For instance, constraining the abundance and configuration of ice giants on orbits exterior to gas giants will significantly bolster our understanding of orbital migration and dynamical instabilities.  Likewise, the radial ordering of systems with different-sized planets at different orbital distances will constrain models of \hbindex{pebble accretion}. 

\section{Summary}

This chapter reviewed the current paradigm of terrestrial planet formation; from dust-coagulation to planetesimal formation to late stage accretion. It discussed the classic scenario of terrestrial planet formation, its well-documented short-comings, and recently proposed alternatives that resolve these issues such as the Grand-Tack model, \hbindex{Early Instability} model and the \hbindex{Planet Formation from Rings} scenario. It also discussed the origins of \hbindex{hot super-Earth}s, placed the Solar System in the context of exoplanets and discussed terrestrial planet formation in exoplanetary systems. The following bullets summarize key advances the study of planet formation, and outstanding questions that continue to drive future investigation:
\begin{itemize}

\item The \hbindex{streaming instability} stands as a promising mechanism for explaining how mm- to cm-sized particles grow to 100 km-scale planetesimals. Yet the \hbindex{streaming instability} require specific conditions to operate. This implies that planetesimals may form in preferential locations (e.g., just beyond the \hbindex{snowline}) that will in turn strongly influence the further formation of the system.

\item Planetesimals grow into planetary embryos (or giant planet cores) by accreting planetesimals or pebbles (or a combination of both). Simulations of planetesimal accretion struggle to grow giant planet cores within the lifetime of \hbindex{protoplanetary disk}s. \hbindex{pebble accretion} may solve this long-standing timescale conflict, but many key aspects of \hbindex{pebble accretion} remain largely unexplored.

\item Three models of the late stage of accretion of terrestrial planet can explain the structure of the inner Solar System: the Grand-Tack, an early giant planet instability and \hbindex{Planet Formation from Rings}. A clear future step in planet formation is to differentiate between these models. Combining \hbindex{N-body simulation}s with geochemical models will undoubtedly be a powerful tool in this pursuit.

\item Of the Solar System's planets, the origin of Mercury is the most mysterious.  In particular, it remains unclear whether the small planet aquired its high iron-content and relatively large core through a giant impact or by forming from a population of chemically distinct planetesimals.  The discovery of additional exoplanets with Mercury-like densities (\hbindex{super-Mercuries}) might help to break degeneracies between these scenarios in the future.

\item  \hbindex{Hot super-Earth}s cannot form by pure in-situ accretion. Super-Earths forming in-situ would grow extremely fast because of the large solid masses required in the inner regions and the corresponding short dynamical timescales. If super-Earths form rapidly in the gaseous disk, they must migrate and not form `in-situ'.

\item Close-in super-Earths may have formed farther from their stars and migrated inward. Migration creates resonant chains anchored at the inner edge of the disk, most of which destabilize when the disk dissipates.  Simulations of this process have been used to quantitatively match the orbital architectures of known super-Earths-hosting systems. No system of super-Earths is likely to have formed in the Solar System simply because it should still exist today (given that disks have inner edges that prevent planets from migrating onto their stars).  

\item The Solar System is quantifiably unusual in its lack of super-Earths and in having a wide-orbit gas giant on a low-eccentricity orbit (a $\lesssim$1\% rarity among Sun-like stars).  These two characteristics may be linked, as Jupiter may have prevented Uranus and Neptune's proto-planet precursors from invading the inner Solar System. Additionally, the fortutious lack of close encounters between Jupiter and Saturn during the Solar System's instability likely prevented the destruction of the terrestrial planets. 

\item Future exoplanet surveys providing data on the occurrence of planets at moderate distances from the host star and more refined constraints on the bulk composition of transiting low-mass planets will shed light on the deep mysteries of terrestrial planet formation.

\end{itemize}

\begin{acknowledgement}

We acknowledge a large community of colleagues whose contributions made this review possible. M.S.C. is supported by NASA Emerging Worlds grant 80NSSC23K0868 and NASA’s CHAMPs team, supported by NASA under Grant No. 80NSSC21K0905 issued through the Interdisciplinary Consortia for Astrobiology Research (ICAR) program. S.N.R. thanks the CNRS's Programme Nationale de Planetologie (PNP) and MITI/80PRIME program for support. The work of R.D. was supported by the NASA Emerging Worlds program, grant 80NSSC21K0387.

\end{acknowledgement}

%  IF you do NOT use bibtex, put comments before the following 2 lines
\bibliographystyle{spbasicHBexo}  %for bibtex
\bibliography{HBexoTemplateBib} %for bibtex-example

\end{document}